\documentclass[aps,preprint,showpacs]{revtex4}
\usepackage{amsmath}
\newcommand{\be}{\begin{equation}}
\newcommand{\ee}{\end{equation}}
\newcommand{\bea}{\begin{eqnarray}}
\newcommand{\eea}{\end{eqnarray}} 
\newcommand{\ba}{\begin{array}}
\newcommand{\ea}{\end{array}}

\newcommand{\bb}{\bibitem}

\begin{document}
%\draft
%\tightenlines

\title{\bf Callan-Symanzik-Lifshitz approach to generic competing systems}
\author{Paulo R. S. Carvalho\footnote{Permanent address: Departamento de 
F\'\i sica, Universidade Federal do Piau\'\i , Campus Ministro Petr\^onio 
Portela, 64049-500, Teresina, PI, Brazil.  e-mail:prscarvalho@ufpi.br} and 
Marcelo M. Leite\footnote{e-mail:mleite@df.ufpe.br}}
\affiliation{{\it Laborat\'orio de F\'\i sica Te\'orica e Computacional, Departamento de F\'\i sica,\\ Universidade Federal de Pernambuco,\\
50670-901, Recife, PE, Brazil}}
%{November 2006}

%\end{center}
\vspace{0.2cm}
\begin{abstract}
{\it We present the Callan-Symanzik-Lifshitz method to approaching the 
critical behaviors of systems with arbitrary competing interactions. Every 
distinct competition subspace in the anisotropic cases define an independent 
set of renormalized vertex parts via normalization conditions with 
nonvanishing distinct masses at zero external momenta. Otherwise, only one 
mass scale is required in the isotropic behaviors. At the critical dimension, 
we prove: i) the existence of the Callan-Symanzik-Lifshitz equations and 
ii) the multiplicative renormalizability of the vertex functions using 
the inductive method. Away from the critical dimension, we utilize the 
orthogonal approximation to compute higher loop Feynman integrals, 
anisotropic as well as isotropic, necessary to get the exponents 
$\eta_{n}$ and $\nu_{n}$ at least up to two-loop level. 
Moreover, we calculate the latter exactly for isotropic behaviors at the same 
perturbative order. Similarly to the computation in the massless formalism, 
the orthogonal approximation is found to be exact at one-loop order. The 
outcome for all critical exponents matches exactly with those computed 
using the zero mass field-theoretic description renormalized at nonvanishing 
external momenta.}
\end{abstract}

\vspace{1cm}

\maketitle

\newpage
\section{Introduction}

Universality classes characterizing the ordinary critical behavior of systems 
undergoing phase transitions \cite{Wilson} have an interesting parallel with 
those associated to competing systems of the Lifshitz type 
\cite{new1,Hornreich}. The ordinary universality hypothesis states that all 
universal amounts characterizing the transition like critical exponents, 
amplitude ratios of certain thermodynamic potentials above and below the 
critical temperature, etc., do not depend on the microscopic details of the 
systems but depend upon the number $N$ of components of the order parameter 
and the space dimension $d$ of the system. Lifshitz universality classes, on 
the other hand, depend on an additional parameter : the number $m$ of spatial 
directions where competition takes place. The simplest competing systems 
belonging to the $m$-axial Lifshitz universality classes correspond to 
$m=1$ (uniaxial) and are examples of complex systems with two ordered 
phases as well as one disordered phase in the vicinity of the Lifshitz 
critical point. The $m$ space directions are called the competition axes. 
Turning off the competing interactions is equivalent to taking the limit 
$m\rightarrow 0$ and the competing system should turn into an ordinary 
critical system.

Lifshitz multicritical 
points appear at the confluence of a disordered phase, 
a uniformly ordered phase and a modulated ordered phase. The spatially 
modulated phase is characterized by a fixed equilibrium wavevector, which 
goes continuously to zero as the Lifshitz point is approached. But how can we 
realize these ideas in terms of concrete models describing actual physical 
systems? The language of magnetic systems is particularly convenient to find 
a simple realization of this critical behavior in terms of a lattice model 
named ANNNI model \cite{new5}. It is a $d$-dimensional Ising model with 
ferromagnetic couplings between first neighbors together with 
antiferromagnetic exchange forces between second neighbors along only one 
space direction. Competition in the interactions arises by varying their 
strengths. More precisely, the Lifshitz point corresponds to a particular 
value of the ratio between the ferromagnetic and antiferromagnetic exchange 
interactions. This situation corresponds to the anisotropic uniaxial 
criticality ($m=1$). The ANNNI model can be applied to describe the critical 
behavior of many systems. However, experiments \cite{becerra1,becerra2,zieba,becerra3} as well as 
theoretical investigations \cite{YCS} have confirmed that the magnetic 
material $MnP$ yields a simple realization of the three-dimensional ANNNI 
model: it has a pure uniaxial $(N=1,d=3,m=1)$ Lifshitz critical behavior. 

The competing axis can be classified according to the number of neighbors 
which are coupled via competing interactions. The ANNNI model can be 
generalized to include $m$ space directions with competing interactions. 
We employ the notation $m_{2} \equiv m$ to label this subspace according to 
the number of neighbors they connect through competing exchange forces. In 
that case the wavevector characterizing the modulated phase has $m_{2}$ 
components. The critical system under study presents an $m_{2}$-fold 
(or $m_{2}$-axial) Lifshitz critical behavior, which can be either 
anisotropic $m_{2}<d$ or isotropic $d=m_{2}$ (close to 8). 

If the field (order parameter) has $N$ components, 
the Lifshitz universality classes are defined by the triplet $(N,d,m_{2})$. 
These universalities correctly reduce to the Ising-like universality class 
$(N,d)$ in the limit $m_{2} \rightarrow 0$ \cite{Leite}. 
These criticalities have encountered applications in many real 
physical systems like liquid crystals \cite{new6}, ferroelectrics \cite{new7}, 
especial polymers \cite{new8}, microemulsions \cite{new9}, 
high-$T_{c}$ superconductors \cite{new10}, magnetic materials \cite{becerra1,becerra2,zieba,becerra3}, 
etc. Furthermore, other aspects have been studied like the formulation of 
quantum phase transitions in Lifshitz points \cite{continenti,ardonne,ghaemi} 
as well as the connection of Lifshitz type field theories with weighted scale 
invariant quantum field theories \cite{Anselmi}. 

From the technical point of view, the original calculation of usual critical 
exponents belonging to the Ising-like universality class (without competition) 
were performed using the renormalization 
group and $\epsilon$-expansion methods via diagrammatic perturbation of 
field-theoretic renormalized massive theories \cite{Wilson}. Some time later 
this method was reformulated such that the former approach was reduced to the 
computation of a few diagrams (1PI vertex parts) yielding the same exponents 
through the use of a renormalized massless scalar field theory at nonvanishing 
external momenta \cite{BLZ,Zinn,new22}. Inspired in this massless framework, 
many calculational schemes in the study of $m$-axial Lifshitz points have been 
put forward \cite{MC,DS,AL}. In particular, the unconventional renormalization 
group arguments in the anisotropic criticalities along with analytical 
solution methods to resolve Feynman path integrals have permitted a better 
comprehension of the critical properties of Lifshitz points using the massless 
scalar field-theoretic setting renormalized at nonvanishing external momenta 
together with the renormalization group equations in the large distance 
infrared regime \cite{preprint,Picture}.  

The method of Refs.\cite{BLZ,Zinn,new22} to obtain critical exponents of 
ordinary critical systems was adapted to include the massive theory 
renormalized at 
zero external momenta. It allowed to understand the connection between the 
infrared behavior of the solution to the renormalization group equation in the 
massless theory with the ultraviolet behavior of the solution to the 
Callan-Symanzik equation \cite{Callan,Symanzik} in the massive theory 
\cite{new25}. 
These calculations are more involved, since the massive Feynman integrals are 
intrinsically more difficult to solve than their counterparts in the massless 
framework. The benefits of this study is that a more direct connection with 
quantum field theory can be made in the ultraviolet regime. Besides, this 
extra information is useful to have a proper understanding of other universal 
quantities like the equation of state, amplitude ratios of certain 
thermodynamic potentials above and below the critical temperature, etc. 
We have shown in a previous work\cite{CL} that it is possible to formulate an 
appropriate renormalized field-theoretic setting for massive scalar fields 
with quartic self interactions in order to compute the critical exponents 
associated to $m$-fold Lifshitz points. There were two main 
steps explicit in that construction. The first one is the appearance of two 
independent mass parameters necessary to describe the two inequivalent space 
(or momentum) directions present in anisotropic criticalities in an 
independent manner. Consequently, the Callan-Symanzik-Lifshitz equations 
allowed to solve the problem at the repulsive ultraviolet fixed points in the 
anisotropic cases with two 
independent mass scales. Second, the isotropic universality class only needs 
one mass scale. Needless to say, the results for the exponents using either 
the massless approach or the massive method are the same \cite{Picture,CL}.

A different generalization of the ANNNI model can be considered to include 
further alternate competing interactions, for instance, up to third neighbors. 
From a phenomenological viewpoint, the first nontrivial example of a higher 
character Lifshitz point (see below) occurs for a uniaxial third character 
Lifshitz point. The three-dimensional phase diagram consists of two parameters 
varying with the temperature, i.e., the ratio of exchange interactions between 
the second and nearest neighbors as well as the ratio of exchange couplings 
between the third and the second neighbors. When the temperature axis is 
projected on the plane of these parameters, there is a region of intersection 
where the different phases associated to the system encounter each other in 
the uniaxial Lifshitz point of third character, whose existence was 
established numerically \cite{Sel}. As before, we should emphasize that the 
uniaxial third character competing axes connect up to third neighbors with 
alternate exchange forces.  When these space dimensions occur along $m_{3}$ 
directions, the critical competing system is said to represent the 
$m_{3}$-axial third character Lifshitz critical 
behavior. These universality classes are characterized by $(N,d,m_{3})$. 
The uniaxial third character 
Lifshitz criticality is the particular case of the $m_{3}$-axial behavior for 
$m_{3}=1$. 

Using these ideas, the generalization of competing systems whose competing 
axes link up to $L$ neighbors via alternate exchange forces can be easily 
understood. Consider a $d$-dimensional Ising model with competition 
interactions connecting $L$ neighbors. Let $m_{L}$ be the competition spatial 
subspace where exchange short range couplings take place with ferromagnetic 
interactions between first neighbors, antiferromagnetic forces between second 
neighbors, ferromagnetic interactions between third neighbors and so on, 
with alternating signs for the exchange forces up to the $L$th neighbors. This 
universality class is characterized by $(N,d,m_{L})$ \cite{new18,new19}. Even 
though the anisotropic $(m_{L}<d)$ and isotropic behaviors $(d=m_{L})$ can 
still be defined, the 
anisotropic situation can be described with only 2 independent scales. Thus, 
this model does not correspond to the most general competing system.

Generic competing systems of the Lifshitz type have been introduced 
recently. A simple realization of their critical properties using the 
language of magnetic systems can be visualized through a lattice model 
called CECI model \cite{generic1,generic2}. It is a generalized Ising model 
with 
several distinct types of competing axes. Each competition subspace is 
perpendicular to each other. It describes a complex system, which in 
a simple situation possesses $L$ ordered phases, as well as one disordered 
phase, near the generic higher character Lifshitz point. This 
is so because the  CECI model contains {\it simultaneously} independent 
competing axes whose exchange couplings are independent in each spatial 
subspace. Consequently, in the anisotropic situation there are 
$m_{2}, m_{3}, ..., m_{L}$ types of competing axes, as well as 
$(d-m_{2}-...-m_{L})$ space directions where there are only 
ferromagnetic couplings among nearest neighbors. We then define 
$m_{1}= d-m_{2}-...-m_{L}$ in order to 
unify the treatment to all subspaces, the competing and noncompeting ones. The 
noncompeting subspace can be viewed as the competing axes with only first 
neighbors interacting ferromagnetically. The ordinary critical behavior can be 
understood in terms of this generic competing system as the isotropic 
particular case for $d=m_{1}$ with $m_{2}= ...=m_{L}=0$. 

The generic $L$-th character Lifshitz anisotropic 
universality classes are now identified by the grid 
$(N,m_{1},m_{2},...,m_{L})$ (actually the same as $(N,d,m_{2},...,m_{L})$), 
therefore generalizing the $L$-th character 
Lifshitz universality class previously discussed, which depends on 
$(N,d,m_{L})$ \cite{new18,new19}. However, can we find physical systems 
which are realizations of higher character Lifshitz points, or does this 
model have only academic interest? For instance, higher character Lifshitz 
points can show up in blends of diblock copolymers. In fact, the earlier 
nomenclature regarding the higher character Lifshitz points can be translated 
into the modern terminology if we identify the $L$-th character Lifshitz point 
with the (former) Lifshitz point of order $(L-1)$\cite{new18}. Thus, it can be 
verified that Lifshitz points of up to 6th character are possible in those 
blends of diblock copolymers \cite{new17}. This is an encouraging evidence 
that competing systems represented by the CECI model might find 
applications in many other examples of actual physical systems yet to be 
discovered. The ANNNI model can be embedded into the CECI 
model, such that the former can be retrieved from the latter when we switch 
off the competing interactions beyond third neighbors by taking 
$m_{2}=1, m_{3}=...=m_{L}=0$. Moreover, the 
isotropic behaviors have their universality classes completely specified by 
$(N,d=m_{L})$, whose critical dimension $d_{c}=4L$ depends as 
well on the number of neighbors coupled via competing interactions. They also 
reduce to the isotropic $m$-axial case when $L=2$. Thus, the usual $m$-axial 
Lifshitz criticality is a particular case of the generic competing system 
described by the CECI model.

In this paper we generalize the framework of renormalized massive scalar 
fields previously introduced to compute critical exponents pertaining to 
the $m$-axial Lifshitz universality classes \cite{CL} in order to study the 
most general competing system. It can be mathematically understood in terms 
of an anisotropic field-theoretic description of a massive renormalized 
$\lambda\phi^{4}$ scalar field theory including arbitrary higher order 
derivatives. Each higher derivative term in 
the Lagrangian density defines a certain type of competing spatial subspace. 
There are $L$ different types of competing axes which result in $L$ sets of 
independent masses and coupling constants. We derive the 
Callan-Symanzik-Lifshitz equations with $L$ independent mass scales, 
therefore generalizing the previous massive formulation for anisotropic 
$m$-axial Lifshitz critical behaviors defined by only two independent mass 
subspaces \cite{CL}. We study the solutions of these generalized 
Callan-Symanzik-Lifshitz equations with several independent mass scales. We 
show that their ultraviolet behavior at the nonattractive ultraviolet fixed 
point is completely equivalent to the solutions of the renormalization group 
equations formulated in the massless theory with several independent momenta 
scales at the infrared fixed point. We focus on renormalized perturbation 
theory in order to compute the critical indices by diagrammatic means within 
this technique. The perfect agreement with those calculated in the massless 
framework at the same loop order is a clear evidence that the universality 
hypothesis is obeyed as expected.     

The critical indices $\eta_{L}$ and $\nu_{L}$ associated to each type of 
competition axes are computed in the anisotropic 
cases using the orthogonal approximation previously introduced in 
\cite{generic1,generic2} to treat the massless field theory formalism. 
We utilize several mass scales, one for each competing subspace, and the 
corresponding renormalized theory is defined at vanishing external momenta. 
Physically, the masses correspond to the independent correlation lengths 
$\xi_{1}$,..., $\xi_{L}$ which go critical simultaneously and naturally take 
place in the anisotropic criticalities. 

In addition, we use a similar framework to define the isotropic behaviors with 
only one mass scale and derive the corresponding Callan-Symanzik equations for 
them. We compute the critical exponents perturbatively using Feynman graph 
techniques. The orthogonal approximation is employed to 
compute the analogous critical exponents in the isotropic behaviors. Besides, 
we shall demonstrate that in spite of the highly nontrivial feature of the 
Feynman integrals in the massive theory, they can be computed exactly giving 
exact results for the critical exponents identical to those found in 
\cite{generic1,generic2}.  

The normalization conditions are presented in Sec.II. There we motivate the 
origin of the several independent mass scales in the anisotropic behaviors 
and show how they can be understood in terms of the phase diagrams of the CECI 
model. A simpler treatment will be given for the isotropic behaviors as well 
with only one mass scale for each type of competing axes.

We discuss the one-loop renormalizability at the critical dimension of the 
various anisotropic and isotropic situations in Sec. III. Using these 
concepts, we prove inductively the finiteness of the multiplicatively 
renormalized vertex parts at all orders in a perturbative expansion and 
demonstrate that the Callan-Symanzik-Lifshitz equations exist in Sec.IV.

In Sec. V we discuss the Callan-Symanzik-Lifshitz (CSL) equations slightly 
away from the critical dimensions ($d=d_{c} - \epsilon_{L}$).We derive those 
equations with several mass scales for the anisotropic criticalities. We 
present the solution of the CSL equations in the ultraviolet regime and 
show that at the ultraviolet nonattractive fixed point it has the same scaling 
form as the solution of the renormalization group equations in the infrared 
regime. We show that the critical exponents 
calculated by diagrammatic means can be identified with the anomalous 
dimension of the field and that of the composite operator at the ultraviolet 
fixed point. 

The computation of the anisotropic critical exponents are presented in Sec.VI. 
The results for Feynman graphs using the orthogonal approximation are derived 
in Appendix A. They are extensively used in Sec.VI in the calculation 
of the critical indices using the orthogonal approximation.

Section VII describes the calculation of the critical exponents for the 
isotropic cases utilizing the orthogonal approximation. The corresponding loop 
integrals are computed in Appendix B. 

Section VIII presents the exact computation of the Feynman integrals in the 
isotropic behaviors. The reason for this explicit computations is 
that the four-point graphs for arbitrary $n$ is quite difficult to get in a 
closed form explicitly. By fixing $n$ we can compute their contribution and 
find a recursion formula for arbitrary $n$. In addition, the two-point 
vertex part graphs are also computed up to three-loop order. They are shown 
to be simpler that the four-point contributions.

Section IX is an exposition of the exact results obtained for generic 
isotropic critical exponents for arbitrary $n$. They are determined 
diagrammatically using the results of Section VIII. They are shown to 
be identical to the exponents previously evaluated using the massless 
formalism. 

We present the discussion of our results along with the conclusions and 
further possible applications of the present method in Sec.X.  

\section{Normalization conditions for the massive theories}

The functional integral representation of the CECI model was first introduced 
in Ref.\cite{generic1}. It corresponds to a $\lambda\phi^{4}$ theory containing 
higher derivatives. The larger the number of neighbors coupled via competing 
interactions, the higher is the power of the derivative terms in the bare 
Lagrangian. In the anisotropic behaviors for generic competing systems, there 
are many simultaneous types of higher derivative terms. The original bare 
Lagrangian density reads
%\begin{subequations}\label{1}
\begin{eqnarray}
L &=& \frac{1}{2}
|\bigtriangledown_{(d- \sum_{n=2}^{L} m_{n})} \phi_0\,|^{2} +
\sum_{n=2}^{L} \frac{\sigma_{n}}{2}
|\bigtriangledown_{m_{n}}^{n} \phi_0\,|^{2} \\ \nonumber
&& + \sum_{n=2}^{L} \delta_{0n}  \frac{1}{2}
|\bigtriangledown_{m_{n}} \phi_0\,|^{2}
+ \sum_{n=3}^{L-1} \sum_{n'=2}^{n-1}\frac{1}{2} \tau_{nn'}
|\bigtriangledown_{m_{n}}^{n'} \phi_0\,|^{2} \\ \nonumber
&&+ \frac{1}{2} \mu_{0}^{2}\phi_0^{2} + \frac{1}{4!}\lambda_0\phi_0^{4} .
\end{eqnarray} 
%\end{subequations}

The first and third summations in the above Lagrangian correspond to the 
effect of the competition occurring in the system. The parameters 
$\sigma_{n}$ and $\tau_{nn'}$ are introduced to guarantee that all terms have 
the same canonical dimension. At the Lifshitz critical region, the fixed 
values of the exchange interactions are transliterated into the conditions 
$\delta_{0n}=\tau_{nn'}=0$ in the above Lagrangian. This simplification 
permits the decoupling of each subspace in Feynman loop integrals, but 
produces the apparent complication that all the higher derivative terms 
become relevant in the free massive propagator. Notice that the 
Lifshitz point is characterized by $\mu_{0}=0$ when the temperature is exactly 
at the Lifshitz value $T=T_{L}$, with $\delta_{0n}=\tau_{nn'}=0$.

It is worthy to separate the momentum subspaces in the form 
$p_{(n)}=(p_{1},..., p_{L})= (\bf{p}, \bf{k_{2}},..., \bf{k_{L}})$ 
where $p_{1} \equiv \bf{p}$ is a vector along $m_{1}$ spatial directions 
connected to the first (non)competing 
subspace, $p_{(2)} \equiv \bf{k_{2}}$ where $\bf{k_{2}}$ is a vector 
along $m_{2}$ space directions associated to the second neighbor competing 
subspace, etc., $p_{(L)} \equiv \bf{k_{L}}$ where $\bf{k_{L}}$ is a vector 
along $m_{L}$ directions associated to the $L$-th neighbor competing 
subspace. The variation of $\kappa_{n}$ in the renormalized theories whose 
starting point is the same bare theory are induced by the existence of 
independent correlation lengths $\xi_{n}$. These $L$ independent flows in the 
momenta can be implemented through $L$ independent renormalization group 
equations for each $m_{n}$-dimensional subspace. Nevertheless, this 
construction is consistent since the apparent overcounting producing 
$L$ independent coupling constants defined in each spatial subset can 
be overcome, for all of them flow to the same infrared nontrivial fixed point.

As discussed in detail in Ref.\cite{generic2} in the massless theory, 
the conditions which define the critical region can be used to perform a 
dimensional redefinition of momentum scales along each type of competing 
subspace as follows. Let $[\bf{p}] = M$ be the mass dimension of the 
quadratic momenta corresponding to the noncompeting subspace $m_{1}$. We can 
get rid of the 
parameter $\sigma_{n}$ appearing in front of each higher derivative term 
characterizing each competing subspace by redefining the associated momenta 
scales through the relations $[\bf{k_{(n)}}] = M^{\frac{1}{n}}$. This 
effectively disentangles each momenta subspace such that they can be treated 
independently. In the anisotropic criticalities there are $L$ subspaces, each 
of them are $m_{L}$-dimensional. The masslessness of the theory at the 
Lifshitz temperature requires that the renormalized 1PI vertex parts (the 
basic objects in this framework related to the thermodynamical potentials of 
the critical system) must be computed at nonzero external momenta in order to 
avoid infrared divergences. For instance, if we consider the vertex functions 
along the $L$-th subspace we use normalization conditions by choosing the 
symmetry point at nonzero external momenta $\kappa_{L}$ along this subspace, 
whereas all the other momenta scales perpendicular to the $m_{L}$ directions 
are set to zero. Therefore, the nonzero momenta used to renormalize the theory 
in each individual subspace can be viewed as a label in parameter space 
defining $L$ independent sets of vertex functions. On the other hand, there 
is no need to consider more than one subspace for isotropic behaviors.    

Now, let us take a look at the simplest CECI model, 
namely, we take $m_{2}=m_{3}=1$ with $m_{n}=0$ for $n>3$. The competition 
are located at the $y$ and $z$ axis. The phase 
diagram (see Fig.1 of Ref.\cite{generic2}) can be described by 
$T, p_{z}=J_{2z}/J_{1z}, p_{1y}=J_{2y}/J_{1y}$ and 
$p_{2y}=J_{3y}/J_{1y}$. We can represent the phase diagram in a very simple 
form through several two-dimensional projections by fixing some parameters and 
varying only two of them. For instance, the diagram $(T, p_{z})$ is associated 
to the second character behavior provided $p_{1y}$ and $p_{2y}$ are kept fixed. On the other hand, the purely third character behavior can be expressed in 
terms of a three-dimensional phase diagram with axes $(T, p_{1y}, p_{2y})$ 
so long as $p_{z}$ is fixed. But this can be further simplified by considering 
only the two-dimensional projection of this phase diagram as suggested above 
if we take instead the variation of $(p_{1y}, p_{2y})$ with $(T, p_{z})$ at 
constant values. Now the superposition of these two two-dimensional phase 
diagrams yields a point of intersection for particular values of the 
temperature which is identified with the generic third character Lifshitz 
point. Notice that the ferromagnetic phase and the two 
modulated phases named $Helical_{2}$ and $Helical_{3}$ (see Fig. 3 of 
Ref.\cite{generic2}) encounter 
themselves at the uniaxial generic third character Lifshitz point. There are 
two first order lines: one of them separates the ordered-$Helical_{2}$ regions 
whereas the other splits the $Helical_{2}-Helical_{3}$ phases. 

This situation 
can be generalized for generic higher character Lifshitz points if we split 
the corresponding multiparameter(multidimensional) phase diagram in 
two-dimensional slices and by superposing them together in a single 
two-dimensional diagram. There are now $L$ modulated phases meeting at the 
$L$-th generic higher character Lifshitz point. We emphasize that each 
competing subspace is defined by its own independent correlation length, 
i.e., $\xi_{1}$ for the subspace with only ferromagnetic exchange forces 
coupling first neighbors, ..., $\xi_{L}$ for the subspace defined by alternate 
signs in the exchange forces up to $L$-th neighbors.

Roughly speaking, we can associate the (inverse of the) mass to the 
correlation length. The anisotropic generic higher character 
universality classes require $L$ independent correlation lengths. In 
close analogy to what has been done for the $m$-axial Lifshitz critical 
behavior (actually a second character Lifshitz behavior) we 
can reexpress the bare Lagrangian density (1) in terms of $L$ independent bare 
masses as follows. 

It is obvious that we can attain the $L$-th generic higher 
character Lifshitz point in the phase diagram outlined above by varying the 
``mass'' coming from the noncompeting subspace. This means that we approach 
the $L$-th generic higher character Lifshitz point from the ferromagnetic 
phase. Simple inspection of the phase diagram indicates that we can reach 
this multicritical point coming from 
{\it any of the several modulated phases}. Then, it is possible to introduce 
$L$ independent bare masses such that they generate the $L$ renormalization 
group flows in parameter space which are compatible with the $L$ correlation 
lengths present in these criticalities. Let us describe the introduction of 
the many independent mass scales beyond these simple phenomenological 
considerations.  

Siegel's method of dimensional reduction suggests in a simple way how to 
introduce mass in ordinary quantum field theories with quadratic derivatives 
in the Lagrangian density\cite{Siegel}. The basic steps for scalar fields 
are: extend the range of the momenta indices and call the extra 
index ``-1'', choose the momentum component associated to this direction 
equal to the mass $p_{-1}=\mu_{0}$ and introduce factors of $i$ to 
re-establish reality $\partial_{1}=ip_{-1}=i \mu_{0}$. The operator $p^{2}$ 
in the higher dimensional space 
(including the extra index ``-1'') results in a massive operator 
$p^{2} + \mu_{0}^{2}$ in $d$ spacetime dimensions. In the Lagrangian (1) the 
metric is Euclidean but the conditions $\delta_{0n}=\tau_{nn'}=0$ can be 
used to perform the dimensional redefinition of the momenta characterizing 
each type of competing axes. The dimensional 
redefinitions turn out to implement independent dilatation invariance along 
the $m_{1}= d - \sum_{n=2}^{L} m_{n}$ noncompeting directions, 
$m_{2}$ subspace, and so on, up to $m_{L}$ space directions. The typical 
momentum combinations which appear in the inverse free propagator is of the 
form $p_{1}^{2} + \sum_{n=2}^{L} (k_{n}^{2})^{n}$.   

The extension of the massive method to include arbitrary types of competing 
axes can be understood from the analysis of $m$-axial critical behavior 
employing massive fields. The introduction of distinct masses using 
Siegel's recipe translates itself in the following conditions: extend the 
range of the vector indices in the $n$ subspace to the ``extra direction'' 
$"-1"(n)$, identify the conjugate momentum component with the mass in 
that subspace $p_{-1(n)}= \mu_{0n}$ and use factors of $i$ to restore 
reality $\partial_{-1(n)}=ip_{-1(n)}= i\mu_{0n}$. If we apply this reduction 
to the $n=1$ subspace, the operator 
$p_{1}^{2} + \sum_{n=2}^{L} (k_{n}^{2})^{n}$ in the higher dimensional space 
turns out to become 
$p_{1}^{2} + \sum_{n=2}^{L} (k_{n}^{2})^{n} + \mu_{01}^{2}$ in $d$ space 
dimensions. When we utilize the same procedure to the subspace $n=2$ with 
$k_{1(2)}=\mu_{02}$, the simplest situation occurs for uniaxial case 
$m_{2}=1$. The combination in the inverse free propagator becomes 
$p_{1}^{2} + k_{2}^{4} + \sum_{n=3}^{L} (k_{n}^{2})^{n}$. If we define the 
internal product in the higher dimensional space with index $``-1(2)"$ by 
$k_{2}^{4} + k_{-1(2)}^{4}$, dimensional reduction yields  $p_{1}^{2} + k_{2}^{4} + \sum_{n=3}^{L} (k_{n}^{2})^{n} \rightarrow p_{1}^{2} + k_{2}^{4} + \sum_{n=3}^{L} (k_{n}^{2})^{n} + \mu_{02}^{4}$. Sticking to simplicity, we can extend 
this procedure to the $n'$ subspace in the uniaxial case $m_{n'}=1$. Choosing 
the internal product in the higher dimensional space including 
the index $-1(n')$ as $k_{n'}^{2n'} + k_{-1(n')}^{2n'}$, the combination 
$p_{1}^{2} + k_{n'}^{2n'} + \sum_{n(\neq n')=2}^{L} (k_{n}^{2})^{n}$ is 
reduced to 
$p_{1}^{2} + k_{n'}^{2n'} + \sum_{n(\neq n')=2}^{L} (k_{n}^{2})^{n} 
+ \mu_{0n'}^{2n'}$. These arguments suggest that it is possible to choose 
the mass defining each competing subspace with the same canonical dimension 
of the momenta along those directions. 

The resulting strategy leads us to 
define the bare masses in the bare Lagrangian density with different powers, 
depending on the chosen subspace we work with. This makes explicit reference 
to the fact that those distinct subspaces are inequivalent. Our experience 
handling labels in the massless theory suggests that when dealing with the 
renormalized theory, the bare mass in each subspace naturally defines a 
renormalized mass and coupling constant characterizing that subspace. We can 
then implement $L$ independent bare masses along with $L$ independent bare 
coupling constants in the bare Lagrangian density. Bearing in mind 
these considerations we write the original bare Lagrangian in the form
\begin{eqnarray}
L &=& \frac{1}{2}
|\bigtriangledown_{(d- \sum_{n=2}^{L} m_{n})} \phi_0\,|^{2} +
\sum_{n=2}^{L} \frac{\sigma_{n}}{2}
|\bigtriangledown_{m_{n}}^{n} \phi_0\,|^{2} \\ \nonumber
&& + \sum_{n=2}^{L} \delta_{0n}  \frac{1}{2}
|\bigtriangledown_{m_{n}} \phi_0\,|^{2}
+ \sum_{n=3}^{L-1} \sum_{n'=2}^{n-1}\frac{1}{2} \tau_{nn'}
|\bigtriangledown_{m_{n}}^{n'} \phi_0\,|^{2} \\ \nonumber
&&+ \frac{1}{2} \mu_{0n}^{2n}\phi_0^{2} + \frac{1}{4!}\lambda_{0n}\phi_0^{4} .
\end{eqnarray} 
We shall focus our attention hereafter in the Lifshitz critical region where 
$\delta_{0n}=\tau_{nn'}=0$ in Eq.(2). The independent bare mass in each 
competing subspace naturally prevents the appearance of infrared divergences 
in calculating Feynman integrals for the associated $1PI$ vertex functions. 
Thus, the label $n$ can be used to define 
independent renormalized $1PI$ vertex functions which have the following 
property: those with arbitrary external momenta $p_{(n)}$ along the $n$-th 
space directions have a nonvanishing bare mass $\mu_{0n}$ and coupling 
constant $\lambda_{0n}$. However, they have vanishing external momenta 
($p_{n'}=0$), bare mass ($\mu_{0n'}=0$) and coupling constant 
($\lambda_{0n'}=0$) along all directions elsewhere (if $n'\neq n$). 
Moreover, the zero 
mass limits of the vertex parts are well-defined and reduce to the cases 
previously investigated in Refs.\cite{generic1,generic2}. 

One important ingredient to complete the description is the existence of $L$ 
independent cutoffs, which are responsible for the independent variations in 
the mass parameters of the anisotropic cases. Each cutoff has the same 
canonical dimension as mass and momenta characterizing the competition 
subspace under consideration. There is a similarity and a difference when we 
compare the massless and the massive approaches. Unlike the massless case, 
the massive theories do not flow to fixed points where they become scale 
invariant: scale invariance is only achieved if the coupling constants in each 
subspace are set exactly at the eigenvalue conditions $u_{n \infty}$, i.e., at 
the non-attractive fixed points. On the other hand, the ultraviolet fixed 
point value of the coupling constants $u_{n \infty}$ are independent of $n$, 
a feature already encountered in our analysis of the massless theory in the 
infrared regime. The isotropic behaviors can be described by the Lagrangian 
(2) with some modifications: the first term does not appear in the isotropic 
situation $d=m_{n}, n=1,2,..,L$ and there is solely one kind of 
bare (renormalized) mass.

We begin by describing the anisotropic renormalization conditions. Except 
for minor modifications, we shall follow the conventions adopted in 
Refs.\cite{generic2,CL}. Let the $n$-th subspace ($n=1,...,L$) be defined by 
a nonvanishing value of bare mass and coupling constant $\mu_{0n} \neq 0, 
\lambda_{0n} \neq 0$ with $\mu_{0n'}=0, \lambda_{0n'} = 0$ for $n'\neq n$. 
The nonvanishing bare parameters induce renormalized parameters $\mu_{n}$ and 
$g_{n}$. The slight change of notation with respect to Ref.\cite{CL} 
in defining the renormalized mass is performed in order to prevent the 
confusion with the number of space directions $m_{n}$ of the $n$-th 
competing subspace. The renormalized $1PI$ vertex parts living in the 
$m_{n}$-dimensional subspace are defined by the following normalization 
conditions:
\begin{subequations}
\begin{eqnarray}
&& \Gamma_{R(n)}^{(2)}(0,\mu_{n}, g_{n}) = \mu_{n}^{2n}, \\
&& \frac{\partial\Gamma_{R(n)}^{(2)}(p_{(n)}, \mu_{n}, g_{n})}{\partial p^{2n}}|_{p_{(n)}^{2n}=0} = 1, \\
&& \Gamma_{R(n)}^{(4)}(0, \mu_{n}, g_{n}) = g_{n}  , \\
&& \Gamma_{R(n)}^{(2,1)}(0, 0;0, \mu_{n}, g_{n}) = 1 .
\end{eqnarray}
\end{subequations} 
In analogy to the massless case, we can fix the renormalized mass scale in 
each competing susbspace by choosing $\mu_{n}^{2n}=1$. 

The isotropic situations can be described directly by using the above 
normalization conditions for the renormalized vertex parts where only one 
type of competing axes are required in that case. Although the scaling 
analysis is a bit different in the isotropic and anisotropic cases, they share 
the same normalization conditions Eq.(3).

\section{One-loop renormalizability at the critical dimension}

In this section we shall investigate the divergence structure of graphs 
corresponding to primitively divergent bare $1PI$ vertex 
parts. In general, the divergences can be expressed in terms of regularized 
expressions which retain their infinite values provided the regulators used 
take appropriate limits. We shall show that these divergences can be expressed 
rather simply in terms of independent cutoffs in the anisotropic cases, when 
there are many independent competing subspaces appearing simultaneously in the 
problem. Otherwise, the isotropic cases only require one type of cutoff. We 
discuss the renormalization of all vertex parts at one-loop level. Many 
results of this section are going to be useful in proving renormalizability 
at arbitrary loop order. We shall restrict our attention throughout only to 
vertex parts which can be renormalized multiplicatively.  

\subsection{Anisotropic Sector}

Consider the noncompeting subspace corresponding to the label $n=1$. The 
vertex parts associated to it possess arbitrary external momenta 
$p_{1}=\bf{p}$ along the $m_{1}= d - \sum_{n=2}^{L} m_{n}$ space directions, 
with nonvanishing bare mass $\mu_{01}$ and coupling constant $\lambda_{01}$. 
The bare primitive divergent vertex parts are $\Gamma_{R(1)}^{(2)}, \Gamma_{R(1)}^{(4)}$ and $\Gamma_{R(1)}^{(2,1)}$. Their perturbative expansions up to 
one-loop order can be written as:
\begin{equation}
\Gamma_{(1)}^{(2)}(p) = p^{2}+\mu_{01}^{2} +\frac{\lambda_{01}}{2} \int \frac{d^{d-\sum_{n=2}^{L} m_{n}}q \Pi_{n=2}^{L} d^{m_{n}}k_{(n)}}{\left(\sum_{n=2}^{L} (k_{(n)}^{2})^{n} + q^{2} + \mu_{01}^{2} \right)}\;\;\;,\\
\end{equation}
\begin{eqnarray}
\Gamma_{(1)}^{(4)}(p_{i}) = && \lambda_{01} - \frac{\lambda_{01}^{2}}{2}
(\int \frac{d^{d-\sum_{n=2}^{L} m_{n}}q \Pi_{n=2}^{L} 
d^{m_{n}}k_{(n)}}{[\sum_{n=2}^{L}(k_{(n)}^{2})^{n} +
(q + p_{1} + p_{2})^{2}+ \mu_{01}^{2}] \left(\sum_{n=2}^{L} (k_{(n)}^{2})^{n} + q^{2} + \mu_{01}^{2} \right)}\;\;\;, \nonumber \\ 
&& \qquad \quad \;\; + (p_{1} \rightarrow p_{3}) 
+ (p_{2} \rightarrow p_{3})),
\end{eqnarray}
\begin{equation}
\Gamma_{(1)}^{(2,1)}(p_{1},p_{2};p_{3}) =  1 - \frac{\lambda_{01}}{2}
\int \frac{d^{d-\sum_{n=2}^{L} m_{n}}q \Pi_{n=2}^{L} 
d^{m_{n}}k_{(n)}}{[\sum_{n=2}^{L}(k_{(n)}^{2})^{n} +
(q + p_{3})^{2}+ \mu_{01}^{2}] \left(\sum_{n=2}^{L} (k_{(n)}^{2})^{n} + q^{2} + \mu_{01}^{2} \right)}.
\end{equation}
It is easy to see that for $m_{2}=...=m_{L}=0$ the above integrals reduce to 
the ordinary noncompeting ones and there is no subintegral involving 
competing momenta. However, we shall keep the values of $m_{n}$ unspecified 
and compute the integrals at the upper critical dimension 
$d_{c}= 4 + \sum_{n=2}^{L}[\frac{(n-1)}{n}]m_{n}$. 

We begin with the computation of the integral $I_{1(1)}$ with a single 
propagator appearing in Eq.(4). We may employ the Schwinger's trick to 
write the propagator in terms of a parametric integral, namely 
\begin{equation}
\frac{1}{{\left(\sum_{n=2}^{L} (k_{(n)}^{2})^{n} + q^{2} + \mu_{01}^{2} \right)}} = \int_{0}^{\infty} d \alpha \;\; 
exp[-\alpha(\sum_{n=2}^{L} (k_{(n)}^{2})^{n} + q^{2} + \mu_{01}^{2})]  .
\end{equation}  
Making use of the identity \cite{Picture}
\begin{equation}
\int_{-\infty}^{\infty} dx_{1}...dx_{m_{n}} exp(-a(x_{1}^{2} + ...+x_{m_{n}}^{2})^{n})
= \frac{1}{2n} \Gamma(\frac{m_{n}}{2n}) a^{\frac{-m_{n}}{2n}} S_{m_{n}},
\end{equation}
we can solve the integrals over the momenta. The resulting expression 
for $I_{1(1)}$ is given by
\begin{equation}
I_{1(1)} = \frac{1}{2} S_{(4-\sum_{n=2}^{L}\frac{m_{n}}{n})} 
\Gamma(2-\sum_{n=2}^{L} \frac{m_{n}}{2n})(\Pi_{n=2}^{L} 
\frac{S_{m_{n}} \Gamma(\frac{m_{n}}{2n})}{2n})
\int_{0}^{\infty} d \alpha \;\; exp[-\alpha \mu_{01}^{2}] \alpha^{-2}  .
\end{equation}
We introduce the cutoff $\Lambda_{1}$ in order to express mathematically the 
ultraviolet divergence implicit in the parametric integral in terms of it. 
The divergence comes from the region for small values of $\alpha$. We then 
regularize the integral by suppressing a domain $(0,\Lambda_{1}^{-2})$ in the 
small $\alpha$ region of integration. It is very simple to introduce in the 
integrand a function of $\alpha$ and $\Lambda_{1}$ $f_{\Lambda_{1}}(\alpha)$ 
which vanishes for $\alpha < \Lambda_{1}^{-2}$ (whose derivatives vanish in the limit $\alpha \rightarrow 0$) and is identically one for 
$\alpha > \Lambda_{1}^{-2}$. The Heaviside function is efficient to produce 
this effect and we choose 
$f_{\Lambda_{1}}(\alpha)= \theta(\alpha - \Lambda_{1}^{-2})$. Integrating by 
parts twice and letting the cutoff go to infinity, the divergence can be 
expressed in terms of the cutoff as 
\begin{equation}
I_{1(1)} = \frac{1}{2} S_{(4-\sum_{n=2}^{L}\frac{m_{n}}{n})} 
\Gamma(2-\sum_{n=2}^{L} \frac{m_{n}}{2n})(\Pi_{n=2}^{L} 
\frac{S_{m_{n}} \Gamma(\frac{m_{n}}{2n})}{2n}) \mu_{01}^{2}(\frac{\Lambda_{1}^{2}}{\mu_{01}^{2}} - ln(\frac{\Lambda_{1}^{2}}{\mu_{01}^{2}}))  .
\end{equation}      
The overall angular factor is different from the pure $\phi^{4}$ field theory 
and appears in the same way in both cases: it shows up whenever a loop 
integral is performed. Proceeding in the standard way, 
it can be absorbed in a redefinition of the coupling constant. Looking at the 
singularity structure of $I_{1(1)}$ we find that its dependence on 
$\Lambda_{1}$ is exactly the same as its counterpart describing 
ordinary critical behavior has in terms of the cutoff, say $\Lambda$, at the 
critical dimension $d=4$ \cite{Zinn,ID}.  

The integral contributing to both $\Gamma_{R(1)}^{(4)}$ and 
$\Gamma_{R(1)}^{(2,1)}$ denoted by $I_{2(1)}$, namely
\begin{equation}  
I_{2(1)}(p) = \int \frac{d^{d-\sum_{n=2}^{L} m_{n}}q \Pi_{n=2}^{L} 
d^{m_{n}}k_{(n)}}{[\sum_{n=2}^{L}(k_{(n)}^{2})^{n} +
(q + p)^{2}+ \mu_{01}^{2}] \left(\sum_{n=2}^{L} (k_{(n)}^{2})^{n} + q^{2} + \mu_{01}^{2} \right)},
\end{equation}
can be performed in a similar fashion. Introduce two Schwinger parameters and 
integrate over the momenta. We are left with two parametric integrals over 
$\alpha_{1}$ and $\alpha_{2}$. Integrate first over $\alpha_{1}$ by defining a 
new variable $\alpha'= \alpha_{1} + \alpha_{2}$. The integration limits of the 
integral over $\alpha'$ turn out to be $(\alpha_{2},\infty)$. This integration 
produces no divergence, but the integral to be done over $\alpha_{2}$ results 
in the ultraviolet divergence for small values of $\alpha_{2}$. We regularize 
this integral exactly as before by introducing the cutoff function 
$f_{\Lambda_{1}}(\alpha_{2})= \theta(\alpha_{2} - \Lambda_{1}^{-2})$ inside 
the integrand. Expanding the integrand in powers of the external momenta, we 
find out that the divergence is present only in the momentum independent term 
and taking the limit $\Lambda_{1} \rightarrow \infty$, we obtain the following 
divergent result
\begin{equation}
I_{2(1)}(p) = \frac{1}{2} S_{(4-\sum_{n=2}^{L}\frac{m_{n}}{n})} 
\Gamma(2-\sum_{n=2}^{L} \frac{m_{n}}{2n})(\Pi_{n=2}^{L} 
\frac{S_{m_{n}} \Gamma(\frac{m_{n}}{2n})}{2n}) ln(\frac{\Lambda_{1}^{2}}{\mu_{01}^{2}})  .
\end{equation}  

Owing to the similarity with the $m$-axial Lifshitz situation, when there is 
only one type of competing axes, we are going to treat all competing subspaces 
at once. We select only the competing axes in the $m_{n}$-dimensional subspace 
where competing interactions couple up to $n$ neighbors. In that case, we 
start with $\mu_{0n'}\equiv \mu_{0n} \delta_{nn'}$, 
$\lambda_{0n'}\equiv \lambda_{0n} \delta_{nn'}$ and external momenta 
$p_{i(n')}\equiv  k'_{i(n)} \delta_{nn'}$. The vertex functions which have 
primitive divergences can be expanded up to one-loop order as
\begin{eqnarray}
&&\Gamma_{(n)}^{(2)}(k'_{(n)}) = (k_{n}^{'2})^{n}+\mu_{0n}^{2n} +\frac{\lambda_{0n}}{2} \int \frac{d^{d-\sum_{n=2}^{L} m_{n}}q \Pi_{n=2}^{L} d^{m_{n}}k_{(n)}}{\left(\sum_{n=2}^{L} (k_{(n)}^{2})^{n} + q^{2} + \mu_{0n}^{2n} \right)}\;\;\;,
\end{eqnarray}
\begin{eqnarray}
&&\Gamma_{(n)}^{(4)}(k'_{i(n)})=  \lambda_{0n} - \frac{\lambda_{0n}^{2}}{2}
(\int \frac{d^{d-\sum_{n=2}^{L} m_{n}}q \Pi_{n=2}^{L} 
d^{m_{n}}k_{(n)}}{[\sum_{(n'\neq n) n'=2}^{L}(k_{(n')}^{2})^{n'} + 
((k_{(n)}+ k^{'}_{1(n)} +k^{'}_{2(n)})^{2})^{n} + q^{2} 
+ \mu_{0n}^{2n}]}\nonumber\\
&&\times \frac{1}{\left(\sum_{n=2}^{L} (k_{(n)}^{2})^{n} + q^{2} + \mu_{0n}^{2n} \right)} \qquad  + (k'_{1(n)} \rightarrow k'_{3(n)}) 
+ (k'_{2(n)} \rightarrow k'_{3(n)})),
\end{eqnarray}
\begin{eqnarray}
&&\Gamma_{(n)}^{(2,1)}(k'_{1(n)},k'_{2(n)};k'_{3(n)}) =  1 - \frac{\lambda_{0n}}{2}
\int \frac{d^{d-\sum_{n=2}^{L} m_{n}}q \Pi_{n=2}^{L} 
d^{m_{n}}k_{(n)}}{[\sum_{(n'\neq n) n'=2}^{L}(k_{(n')}^{2})^{n'} + 
((k_{(n)}+k'_{3(n)})^{2})^{n}  + q^{2}+ \mu_{0n}^{2n}]} \nonumber\\ 
&&\times \frac{1}{\left(\sum_{n=2}^{L} (k_{(n)}^{2})^{n} + q^{2} + \mu_{0n}^{2n} \right)}.
\end{eqnarray}

In order to perform these integrals, we recall that the bare mass parameter 
has $\frac{1}{n}$ of the canonical dimension of the momenta along the 
noncompeting (quadratic) subspace. We then choose the cutoff $\Lambda_{n}$ 
associated to this subspace with the same canonical dimension as $\mu_{0n}$. 
When using the Schwinger parameters to perform the integrals over the momenta 
variables, we encounter the divergences implicit over the parametric integrals 
in the region of small values of $\alpha$. We cutoff the parametric integrals 
introducing in the integrand the regularization function 
$f_{\Lambda_{n}}(\alpha)= \theta(\alpha - \Lambda_{n}^{-2n})$, which 
restricts the integration domain to the 
$\Lambda_{n}^{-2n}\leq \alpha \leq \infty$. Denote the integrals contributing 
to the two-point function at one-loop by $I_{1(n)}$ whereas  $I_{2(n)}$ 
represent the contributions to $\Gamma_{(n)}^{(4)}$ 
(and $\Gamma_{(n)}^{(2,1)}$), respectively. Taking the limit 
$\Lambda_{n} \rightarrow \infty$ we get to the following expressions for these 
objects $(n=2,...,L)$
\begin{equation}
I_{1(n)} = \frac{1}{2} S_{(4-\sum_{n=2}^{L}\frac{m_{n}}{n})} 
\Gamma(2-\sum_{n=2}^{L} \frac{m_{n}}{2n})(\Pi_{n=2}^{L} 
\frac{S_{m_{n}} \Gamma(\frac{m_{n}}{2n})}{2n}) \mu_{0n}^{2n}(\frac{\Lambda_{n}^{2n}}{\mu_{0n}^{2n}} - ln(\frac{\Lambda_{n}^{2n}}{\mu_{0n}^{2n}}))  ,
\end{equation}    
\begin{equation}
I_{2(n)}(k'_{(n)} = \frac{1}{2} S_{(4-\sum_{n=2}^{L}\frac{m_{n}}{n})} 
\Gamma(2-\sum_{n=2}^{L} \frac{m_{n}}{2n})(\Pi_{n=2}^{L} 
\frac{S_{m_{n}} \Gamma(\frac{m_{n}}{2n})}{2n}) ln(\frac{\Lambda_{n}^{2n}}{\mu_{0n}^{2n}})  .
\end{equation} 
Comparing the above equations with their counterparts in the noncompeting 
(quadratic) subspace, it is obvious that we can now unify all subspaces by 
taking $n=1,...,L$ in the above formulae. This unification 
will be useful in our description of the renormalization of masses and 
coupling constants at one-loop order. The fact of the matter is that when we 
use the normalization conditions at zero external momenta in each subspace 
with a set of finite renormalized parameters ($\mu_{n}$, $g_{n}$) 
starting from infinite bare quantities ($\mu_{0n}$, $\lambda_{0n}$), we can 
express the former in terms of the latter via the following equations: 
\begin{equation}
\mu_{n}^{2n} = \mu_{0n} + \frac{\lambda_{0n}}{2} I_{1(n)}(0),
\end{equation}
\begin{equation}
g_{n} = \lambda_{0n} - \frac{3 \lambda_{0n}^{2}}{2} I_{2(n)}(0).
\end{equation}
If we go the other way around by writing the bare parameters in terms of the 
renormalized ones and getting rid of higher order corrections in the 
renormalized coupling constants, these simultaneous operations render the 
original bare vertex parts $\Gamma_{(n)}^{(2)}$ and $\Gamma_{(n)}^{(4)}$ 
finite. In fact, the corresponding finite vertex parts are given by 
$(n=1,...,L)$
\begin{equation}
\Gamma_{(n)}^{(2)}(p_{(n)})= (p_{(n)})^{2n} + \mu_{n}^{2n},
\end{equation}
\begin{equation}
\Gamma_{(n)}^{(4)}(p_{i(n)}) = g_{n} -\frac{g_{n}^{2}}{2}
(I_{2(n)}(p_{1(n)} + p_{2(n)}) + I_{2(n)}(p_{1(n)} + p_{3(n)}) 
+ I_{2(n)}(p_{2(n)} + p_{3(n)}) - 3I_{2(n)}(0)).
\end{equation} 

The bare vertex parts $\Gamma_{(n)}^{(N)}$ with $N>4$ have skeleton 
expansions. There are $L$ independent sets of skeleton expansions, one set 
for each competing subspace. Thus, their diagrammatic expansions result in 
finite expressions at two-loop level when they are written in terms of the 
renormalized mass(es) and coupling constant(s) at one-loop order.

In order to study insertion of composite operators and their renormalization, 
we analyze the bare vertex $\Gamma_{(n)}^{(2,1)}$. They are not automatically 
finite by the reparametrizations turning the bare mass and coupling constant 
into those renormalized amounts. We can define the renormalized (finite) 
vertex part $\Gamma_{R(n)}^{(2,1)}$ as
\begin{equation}
\Gamma_{R(n)}^{(2,1)}(p_{1(n)}, p_{2(n)}; p_{3(n)}; g_{n},\mu_{n}) = Z_{\phi^{2}(n)} 
\Gamma_{(n)}^{(2,1)}(p_{1(n)}, p_{2(n)}; p_{3(n)}; \lambda_{0n},\mu_{0n}, \Lambda_{n}). 
\end{equation} 
The normalization conditions (3d) require that, at this order in the loop 
expansion, the normalization functions are given by
\begin{equation}
Z_{\phi^{2}(n)} = 1 + \frac{g_{n}}{2} I_{2(n)}(0).
\end{equation}
Replacing this expression into the definition of $\Gamma_{R(n)}^{(2,1)}$ turns 
out to make this vertex function finite, which may be written as
\begin{equation}
\Gamma_{R(n)}^{(2,1)}(p_{1(n)}, p_{2(n)}; p_{3(n)}; g_{n},\mu_{n}) =  
1 - \frac{g_{n}}{2} (I_{2(n)}(p_{3(n)}) - I_{2(n)}(0)) .
\end{equation}  
In most applications we shall describe insertion at zero external momenta, 
i.e., we refer to the vertex function 
$\Gamma_{R(n)}^{(2,1)}(p_{(n)}, - p_{(n)}; 0; g_{n},\mu_{n})$. In this manner, we can 
address the multiplicative renormalizability and obtain independent flows in 
the parameter spaces of vertex parts including arbitrary types of composite 
operators, irrespective of the competing subspace under scrutiny.

The remaining multiplicatively renormalized vertex parts including composite 
operators $\Gamma_{R(n)}^{(N,L)}$ with $(N,L)>(2,1)$ have skeleton expansions. 
Consequently, they are finite at two-loop level whenever we use 
$g_{n}, \mu_{n}$ and $\Gamma_{R(n)}^{(2,1)}(p_{1(n)}, p_{2(n)}; p_{3(n)}; g_{n},\mu_{n})$ (or $Z_{\phi^{2}(n)}$) inside their one-loop subgraphs, due to the 
following high momentum pattern of the primitively divergent vertex functions
\begin{equation}
|\Gamma_{R(n)}^{(2)}(\rho_{n} p_{(n)})| \leq \rho_{n}^{2n}\; \times 
\;\;power\;\;of\;\; ln\rho_{n}, \nonumber
\end{equation}
\begin{equation}
|\Gamma_{R(n)}^{(4)}(\rho_{n} p_{i(n)})| \leq  \;\; power\;\; of\;\; 
ln\rho_{n}, \nonumber
\end{equation}  
\begin{equation}
|\Gamma_{R(n)}^{(2,1)}(\rho_{n} p_{1(n)},\rho_{n} p_{1(n)};
\rho_{n} q_{(n)})  | \leq  \;\;power\;\; of\;\; ln\rho_{n}, \nonumber
\end{equation}  
at every finite order in the limit $\rho_{n} \rightarrow \infty$. This large 
momentum behavior is not going to be derived here, but we shall assume that it 
is valid henceforth. In other words, the Born values of the various 
renormalized vertex functions are modified only by powers of logarithms. These 
modifications are caused by the interactions in every perturbative order at 
the loop expansion, and we can use a power counting reasoning to approach the 
multiplicative renormalizability at the critical dimension to all loop orders. 
This (nonperturbative) proof of multiplicative renormalizability corresponding 
to the Callan-Symanzik-Lifshitz massive method shall be analyzed after our 
construction of similar renormalization background for isotropic behaviors at 
one-loop order, to which we turn our attention next. 

\subsection{Isotropic Sector}

Recall that for arbitrary higher character isotropic Lifshitz critical 
behaviors there is only one type of competing subspace. The critical dimension 
of the isotropic critical behaviors when there are $n$ neighbors along each 
space dimension interacting via alternate exchange forces is $d=m=4n$, with 
$n=1,...,L$. The one-loop required vertex parts which are primitively 
divergent are given by
\begin{equation}
\Gamma_{(n)}^{(2)}(k) = ((k)^{2})^{n}+\mu_{0n}^{2n} +\frac{\lambda_{0n}}{2} 
\int \frac{d^{4n}k'}{[\bigl((k')^{2}\bigr)^{n} + \mu_{0n}^{2n}]}\;\;\;,\\
\end{equation}
\begin{equation}
\Gamma_{(n)}^{(4)}(k_{i}) = \lambda_{0n} - \frac{\lambda_{0n}^{2}}{2}(\int \frac{d^{4n}k'}{[\bigl((k' + k_{1} + k_{2})^{2}\bigr)^{n} + \mu_{0n}^{2n}]
[((k')^{2})^{n} + \mu_{0n}^{2n}]}) + (k_{1} \rightarrow k_{3}) + (k_{2} \rightarrow k_{3}))\;\;\;,\\
\end{equation}
\begin{equation}
\Gamma_{(n)}^{(2,1)}(k_{1},k_{2};k_{3}) =  1 - \frac{\lambda_{0n}}{2} \int \frac{d^{4n}k'}{[\bigl((k' + k_{3})^{2}\bigr)^{n} + \mu_{0n}^{2n}][((k')^{2})^{n} + \mu_{0n}^{2n}] }\;\;\;.
\end{equation}   
Let $I_{1(n)}$ and $I_{2(n)}$ be the one-loop integrals involved in the 
calculation of $\Gamma_{(n)}^{(2)}$ and  $\Gamma_{(n)}^{(4)}$, respectively. 
The computation of the integrals are entirely analogous to the $nth$ 
competing subspace of the anisotropic sector: the cutoff $\Lambda_{n}$ has 
the same canonical dimension of $\mu_{0n}$ and the removal of small values 
of the parametric integrals in the Schwinger parameters are implemented via 
Heaviside's function. The only difference with respect to the 
anisotropic case study in the $nth$ competing subspace is that the angular 
factor is different. We then obtain
\begin{equation}  
I_{1(n)}(k) = \frac{1}{2n}\;\;S_{4n}\;\; 
\mu_{0n}^{2n}(\frac{\Lambda_{n}^{2n}}{\mu_{0n}^{2n}} 
- ln(\frac{\Lambda_{n}^{2n}}{\mu_{0n}^{2n}})), 
\end{equation} 
\begin{equation}
I_{2(n)}(k)= \frac{1}{2n}\;\;S_{4n}
\;\;ln(\frac{\Lambda_{n}^{2n}}{\mu_{0n}^{2n}})  .
\end{equation} 
Renormalization of the mass and coupling constant can be accomplished through 
the utilization of the normalization conditions (3a) and 
(3c). When expressed as functions of the bare amounts (which have infinite 
values) they take the form
\begin{equation}
\mu_{n}^{2n}= \mu_{0n} + \frac{\lambda_{0n}}{2}I_{1(n)}(0),
\end{equation}
\begin{equation}
g_{n} = \lambda_{0n} - \frac{3 \lambda_{0n}^{2}}{2}I_{2(n)}(0).
\end{equation}
We could have written the above expressions in the opposite direction. In 
that case, either the bare functions 
$\Gamma_{(n)}^{(N)}(k_{i})(\mu_{n}, g_{n})$ can be rendered convergent at 
one-loop order provided $N \leq 4$, or they are skeleton expansions. To 
conclude, whenever we use the normalization condition (3d) along with the 
appropriate version of Eq. (22), which amounts to define the normalization 
function $Z_{\phi^{2} (n)}$, they produce a convergent expression for the 
renormalized vertex $\Gamma_{R(n)}^{(2,1)}(k_{1},k_{2};k_{3})$. Consequently, 
this leads to $Z_{\phi^{2} (n)}= 1 + \frac{g_{n}}{2} I_{2(n)}(0)$. 
\par Next, let us tackle the issue of multiplicative renormalization 
in order to extend the one-loop discussion just explained in the present 
section to arbitrary loop order employing the Callan-Symanzik-Lifshitz 
equations. 

\section{The Callan-Symanzik-Lifshitz equations at the critical dimensions}
\par Let us establish the inductive proof of multiplicative renormalizability 
and prove the existence of the Callan-Symanzik-Lifshitz equation in the 
presence of various simultaneous types of competing axes as well as when 
solely one type of competing axes take place in the generalized Lifshitz 
critical behaviors. 

Although we shall not make explicit reference, we shall be 
concerned with vertex parts which are regularized using cutoffs as described 
in the previous section. It is implicitly assumed that this is the case 
in the remainder of the present section. Our discussion here is restricted 
to the critical dimension of the theory.

\subsection{Anisotropic behaviors}
\par We are going to study the physical system at its upper critical dimension 
$d_{c}= 4 + \sum_{n=2}^{L}[\frac{(n-1)}{n}]m_{n}$. The multiplicatively 
renormalized vertex parts including composite operators are defined by
\begin{eqnarray}
\Gamma_{R(n)}^{(N,L)} (p_{i (n)}, Q_{i(n)}, g_{n}, \mu_{n})
&=& Z_{\phi (n)}^{\frac{N}{2}} Z_{\phi^{2} (n)}^{L}
\Gamma^{(N,L)} (p_{i (n)}, Q_{i (n)}, \lambda_{0n}, \mu_{0n}, \Lambda_{n}).
\end{eqnarray}
We emphasize that vertex function with $(N,L)=(0,2)$ which are 
additively renormalized are precluded from our analysis. In the above 
expression, $p_{i (n)}$ ($i=1,...,N$) are the external momenta associated to
the $N$ external legs of $\phi$ operators, $Q_{i (n)}$ ($i=1,...,L$) are the 
external momenta corresponding to the
$L$ insertions of $\phi^{2}$ composite operators and the independent cutoffs 
$\Lambda_{n}$ are required to depict the regularization process in each 
inequivalent subspace. In the present section, we shall suppose that every 
bare diagram at any loop order can be implicitly regularized with the 
appropriate cutoff.
\par When we apply the operation $\frac{\partial}{\partial \mu_{0n}^{2n}}$ to 
the bare vertex part $\Gamma_{(n)}^{(N,L)} (p_{i(n)}, Q_{i(n)},\lambda_{0n}, \mu_{0n}, \Lambda_{n})$ with fixed $\lambda_{0n}$ and 
$\Lambda_{n}$, we obtain a zero momentum insertion of the operator $\phi^{2}$ 
in the vertex part $\Gamma_{(n)}^{(N,L)} (p_{i (n)}, Q_{i (n)}, \lambda_{0n}, \mu_{0n}, \Lambda_{n})$, namely     
$\Gamma_{(n)}^{(N,L+1)} (p_{i (n)}, Q_{i (n)}, 0, \lambda_{0n}, \mu_{0n}, \Lambda_{n})$. Then, we have
\begin{equation}
\frac{\partial}{\partial \mu_{0n}^{2n}}\Gamma_{(n)}^{(N,L)}(p_{i (n)}, Q_{i (n)}, \lambda_{0n}, \mu_{0n}, \Lambda_{n})= 
\Gamma_{(n)}^{(N,L+1)} (p_{i (n)}, Q_{i (n)}, 0, \lambda_{0n}, \mu_{0n}, \Lambda_{n}) .
\end{equation}   
\par Eq.(32) can be employed to transform the bare quantities into the 
renormalized amounts. Writing the bare mass $\mu_{0n}$ representing the 
$m_{n}$-dimensional subspace in terms of the renormalized mass 
and coupling constant ($\mu_{0n}= \mu_{0n}(\mu_{n},g_{n})$) and applying the 
chain rule, the renormalized vertex functions 
obey the equation
\begin{eqnarray}
&&(2n \rho_{n} \frac{\partial}{\partial \mu_{n}^{2n}} +
\frac{\alpha_{n}}{m_{n}^{2n}} \frac{\partial}{\partial g_{n}}
- \frac{1}{2} N \frac{\kappa_{n}}{\mu_{n}^{2n}} - L \frac{\pi_{n}}{\mu_{n}^{2n}})
\Gamma_{R(n)}^{(N,L)} (p_{i (n)}, Q_{i (n)}, g_{n},
\mu_{n})= \\ \nonumber
&& \Gamma_{R(n)}^{(N,L+1)} (p_{i (n)}, Q_{i (n)},0, g_{n},\mu_{n}) ,
\end{eqnarray}  
where $2n \rho_{n} 
= \frac{\partial \mu_{n}^{2n}}{\partial \mu_{0n}^{2n}} Z_{\phi^{2}(n)}$, 
$\frac{\alpha_{n}}{\mu_{n}^{2n}}=  
Z_{\phi^{2}(n)} \frac{\partial g_{n}}{\partial \mu_{0n}^{2n}}$, 
$\frac{\kappa_{n}}{\mu_{n}^{2n}}=  Z_{\phi^{2}(n)} \frac{\partial lnZ_{\phi(n)}}{\partial \mu_{0n}^{2n}}$, 
$\frac{\pi_{n}}{\mu_{n}^{2n}}=  Z_{\phi^{2}(n)} \frac{\partial lnZ_{\phi^{2}(n)}}{\partial \mu_{0n}^{2n}}$. If we define the functions 
$\beta_{n} (= \frac{\alpha_{n}}{\rho_{n}})
= \mu_{n} \frac{\partial g_{n}}{\partial \mu_{n}}$, 
$\gamma_{\phi(n)}(= \frac{\kappa_{n}}{\rho_{n}})
= \mu_{n} \frac{\partial ln Z_{\phi (n)}}{\partial \mu_{n}}$ and 
$\gamma_{\phi^{2}(n)}(= - \frac{\pi_{n}}{\rho_{n}})
= - \mu_{n} \frac{\partial ln Z_{\phi^{2} (n)}}{\partial \mu_{n}}$, together
 with the multiplication of the last equation by $\frac{\mu_{n}^{2n}}{\rho_{n}}$ yield the result
\begin{eqnarray}
&&(\mu_{n} \frac{\partial}{\partial \mu_{n}} + 
\beta_{n} \frac{\partial}{\partial g_{n}}
- \frac{N}{2} \gamma_{\phi(n)} + L \gamma_{\phi^{2}(n)})
\Gamma_{R(n)}^{(N,L)} (p_{i (n)}, Q_{i (n)}, g_{n},
\mu_{n})= \\ \nonumber
&& 2n \mu_{n}^{2n} \frac{\partial \mu_{0n}^{2n}}{\partial \mu_{n}^{2n}} Z_{\phi^{2}(n)}^{-1} \Gamma_{R(n)}^{(N,L+1)} (p_{i (n)}, 
Q_{i (n)};0, g_{n}, \mu_{n})  \;\;.
\end{eqnarray} 
\par In order to express last equation purely in terms of renormalized objects 
we make use of  the normalization conditions 
$\Gamma_{R(n)}^{(2)}(0)= \mu_{n}^{2n}$,
$\Gamma_{R(n)}^{(2,1)}(0)= 1$ for the particular case $(N,L)=(2,0)$. This 
results in the Callan-Symanzik-Lifshitz equations (CSLE) for anisotropic 
Lifshitz points of generic competing systems:  
\begin{eqnarray}
&&(\mu_{n} \frac{\partial}{\partial \mu_{n}} + 
\beta_{n} \frac{\partial}{\partial g_{n}}
- \frac{N}{2} \gamma_{\phi(n)} + L \gamma_{\phi^{2}(n)})
\Gamma_{R(n)}^{(N,L)} (p_{i (n)}, Q_{i (n)}, g_{n},
\mu_{n})= \\ \nonumber
&& \mu_{n}^{2n}(2n - \gamma_{\phi (n)}) \Gamma_{R(n)}^{(N,L+1)} (p_{i (n)}, 
Q_{i (n)};0, g_{n}, \mu_{n})  \;\;.
\end{eqnarray} 
\par Some comments are in order \cite{Zinn}. Initially, apply a derivative 
$\frac{\partial}{\partial p_{(\tau)}^{2\tau}}$ of the CSLE  
at zero external momenta for $(N,L)=(2,0)$ and use the normalization 
conditions of section II. This means
\begin{equation}
-\gamma_{\phi (n)} = \mu_{n}^{2n}(2n - \gamma_{\phi (n)})
\frac{\partial}{\partial p_{(n)}^{2n}}\Gamma_{R(n)}^{(2,1)} (p_{(n)},- p_{(n)};0, g_{n}, \mu_{n})|_{p_{(n)}^{2n}=0}.
\end{equation}
At one-loop order we already encountered from Eq.(24) that 
$\Gamma_{R(n)}^{(2,1)} (p_{(n)},- p_{(n)};0, g_{n}, \mu_{n})= 1 
+ O(g_{n}^{2})$. It is easy to see that $\gamma_{\phi (n)}$ begins at 
$O(g_{n}^{2})$. Next, substitute $(N,L)=(4,0)$ in the CSLE. Taking 
advantage of  the normalization conditions leads to
\begin{equation}
 \beta_{n} - 2\gamma_{\phi (n)} g_{n} = \mu_{n}^{2n}(2n - \gamma_{\phi (n)})\Gamma_{R(n)}^{(4,1)} (0,0,0,0;0, \mu_{n}, g_{n})\;\;\;.
\end{equation}
The first contribution to $\Gamma_{R(n)}^{(4,1)} (0,0,0,0;0, \mu_{n}, g_{n})$ 
is $O(g_{n}^{2})$. From last equation we discover that 
$\beta_{n}$ is $O(g_{n}^{2})$. Consequently, the operation 
$\beta_{n} \frac{\partial}{\partial g_{n}}$ is $O(g_{n})$, differently from 
the operation $\mu_{n}\frac{\partial}{\partial \mu_{n}}$ 
which is $O(g_{n}^{0})$. Now, taking $N=2$ and $L=1$ into the CSLE along with 
the normalization condition 
$ \Gamma_{R(n)}^{(2,1)} (0,0;0, g_{n}, \mu_{n})=1$, we find
\begin{equation}
-\gamma_{\phi (n)} + \gamma_{\phi^{2}(n)} = \mu_{n}^{2n}(2n - \gamma_{\phi (n)})\Gamma_{R(n)}^{(2,2)} (0,0;0,0, \mu_{n}, g_{n})\;\;\;.
\end{equation}
As $\Gamma_{R(n)}^{(2,2)}$ starts at $O(g_{n})$, we find out
directly that $\gamma_{\phi^{2} (n)} = O(g_{n})$. We have at hand the 
basic requirements to pursuing the inductive proof of multiplicative 
renormalizability at all orders of perturbation theory. We turn now our 
attention to this subject.
\par We begin the inductive proof of multiplicative renormalizability with the 
hypothesis that the renormalized vertices defined by Eq.(32) have been 
transformed into convergent expressions up to the $Lth$-loop order at fixed 
$\mu_{n}$ and $g_{n}$ in the limit $\Lambda_{n} \rightarrow \infty$. Thus, 
we commence with the claim that in the infinite cutoff limit 
$\Gamma_{R(n)}^{(2)}$, $\Gamma_{R(n)}^{(4)}$ and $\Gamma_{R(n)}^{(2,1)}$ 
are finite at order $g_{n}^{L}$, $g_{n}^{L+1}$ and $g_{n}^{L}$, respectively. 
We suppress the arguments of the renormalized vertex functions in order to  
simplify the succeeding discussion. 
\par The preliminary observations considered above at one-loop can be 
generalized to $L$-loop level. Using Eq.(37) at this order in the 
coupling constant $g_{n}$, we find out that 
$\gamma_{\phi (n)}$ is finite at $O(g_{n}^{L})$. From Eq.(38) at 
$(L+1)$-loop order, $\Gamma_{R(n)}^{(4,1)}$ in the right-hand side (rhs) is 
already finite at $(L+1)$-loop order for it has a skeleton expansion. 
On the other hand, at $(L+1)$-loop order this vertex part is of order 
$g_{n}^{L+2}$. In other words, the combination 
$(\beta_{n} - 2\gamma_{\phi(n)} g_{n})$ is convergent at 
$O(g_{n}^{L+2})$. As $\gamma_{\phi (n)}$ is finite at 
$O(g_{n}^{L})$, this implies that $\beta_{n}$ is finite at 
$O(g_{n}^{L+1})$.
\par We can write the CSLE alternatively as
\begin{equation}
(\mu_{n} \frac{\partial\Gamma_{R(n)}^{(N,M)}}{\partial m_{n}}) = 
(-\beta_{n} \frac{\partial}{\partial g_{n}}
+ \frac{N}{2} \gamma_{\phi(n)} - M \gamma_{\phi^{2}(n)})
\Gamma_{R(n)}^{(N,M)} + \mu_{n}^{2n}(2n - \gamma_{\phi (n)}) \Gamma_{R(n)}^{(N,M+1)}   \;\;.
\end{equation}    
Our aim is to prove that we can find a finite result for the rhs of this 
equation at $(L+1)$-loop order. Note that 
$\Gamma_{R(n)}^{(N,M)}$ in the rhs is only needed at $Lth$ loop order, since 
its coefficients $\beta_{n} \frac{\partial}{\partial g_{n}}$ and $\gamma_{\phi^{2}(n)}$ are at least of order $g_{n}$. In 
the last piece of the rhs, either $\Gamma_{R(n)}^{(N,M+1)}$ has a skeleton 
expansion (convergent at $(L+1)$-loop order) or the CSLE must be 
iterated.
\par Let us fix our attention in the case $N=4, M=0$ in Eq.(40). Both terms 
in the rhs are finite at ($L+1)$-loop order. We then conclude that 
$(\mu_{n} \frac{\partial\Gamma_{R(n)}^{(4)}}{\partial \mu_{n}})$ is also 
convergent at this order. 
\par The proof that $(\mu_{n} \frac{\partial\Gamma_{R(n)}^{(4)}}{\partial \mu_{n}})$ is finite at $(L+1)$-loop order as well can be achieved by considering 
the perturbative integration of $\Gamma_{R(n)}^{(4)}$. At the critical 
dimension $\Gamma_{R(n)}^{(4)}$ is dimensionless; it gets unchanged under 
independent dilatations in their dimensionful parameters:
\begin{equation}
(p_{n}, \mu_{n}, \Lambda_{n}, g_{n}) \rightarrow 
(\rho_{n} p_{n}, \rho_{n} m_{n}, \rho_{n} \Lambda_{n}, g_{n}).
\end{equation}
The choice $\rho_{n} = \frac{1}{\mu_{n}}$ together with the dilatation 
invariance stated above can be written as
\begin{equation}
\Gamma_{R(n)}^{(4)}(p_{n}, \mu_{n}, \Lambda_{n}, g_{n}) = 
\Gamma_{R(n)}^{(4)}(\frac{p_{n}}{\mu_{n}}, \frac{\Lambda_{n}}{\mu_{n}}, g_{n}).
\end{equation}
Taking $N=4, M=0$ in Eq.(40) for arbitrary 
external momenta at order $g_{n}^{L+2}$, it can be reexpressed in the 
following manner in terms of a running variable $\mu'_{n}$
\begin{equation}
(\mu'_{n} \frac{\partial\Gamma_{R(n)}^{(4)}}{\partial \mu'_{n}}(\frac{p_{n}}
{\mu'_{n}}, \frac{\Lambda_{n}}{\mu'_{n}}, g_{n}))|_{L+2} = 
f_{(n)}^{(4)}(\frac{p_{n}}{\mu'_{n}}, \frac{\Lambda_{n}}{\mu'_{n}}, g_{n}))|_{L+2} .
\end{equation}
\par First, write the running mass as a function of a dimensionless variable 
$\mu'_{n}= \frac{\mu_{n}}{\alpha}$. The integration of $\mu'_{n}$ in the 
interval $[\infty, \mu_{n}]$ amounts to integrate over $\alpha$ in the region 
$(0,1)$. We then employ the normalization conditions 
(3c) at the limit $\mu'_{n}= \infty$ as boundary conditions. We encounter 
the following result 
\begin{equation}
[\Gamma_{R(n)}^{(4)}
(\frac{p_{n}}{\mu_{n}}, \frac{\Lambda_{n}}{\mu_{n}}, g_{n})]|_{L+2} 
= g_{n} - \int_{0}^{1} \frac{d \alpha}{\alpha} [f_{(n)}^{(4)}(\alpha 
\frac{p_{n}}{\mu_{n}}, \alpha \frac{\Lambda_{n}}{\mu_{n}}, g_{n})]|_{L+2} .
\end{equation}
\par From  what we have been discussing, $[f_{(n)}^{(4)}]|_{L+2}$ can be 
written  entirely in terms of lower order amounts which, by assumption, all 
possess a finite value at $\Lambda_{n} \rightarrow \infty$. Since 
$[f_{(n)}^{(4)}]|_{L+2}$ is analytic for small momenta, the limit of zero 
external momenta can be taken safely. Indeed, this does not introduce any 
difficulty into the integral over $\alpha$. Hence, we are led to 
\begin{equation}
[\Gamma_{R(n)}^{(4)}
(\frac{p_{n}}{\mu_{n}}, \infty, g_{n})]|_{L+2} 
= g_{n} - \int_{0}^{1} \frac{d \alpha}{\alpha} f_{(n)}^{(4)}(\alpha 
\frac{p_{n}}{\mu_{n}}, \infty, g_{n})|_{L+2} ,
\end{equation}
which demonstrates that the renormalized vertex $\Gamma_{R(n)}^{(4)}$ can be 
successfully related to a finite integral over lower order renormalized 
vertex functions. 
\par Considering Eq.(40) for the values $N=2,M=1$ at order 
$g_{n}^{L+1}$ yields
\begin{equation}
(\mu_{n} \frac{\partial\Gamma_{R(n)}^{(2,1)}}{\partial \mu_{n}})|_{L+1} = 
[-\beta_{n} \frac{\partial}{\partial g_{n}} 
\Gamma_{R(n)}^{(2,1)}]|_{L+1}
+ [(\gamma_{\phi(n)} - \gamma_{\phi^{2}(n)})
\Gamma_{R(n)}^{(2,1)}]|_{L+1} + \mu_{n}^{2 \tau}[(2n - \gamma_{\phi (n)}) \Gamma_{R(n)}^{(2,2)}]|_{L+1}   \;\;.
\end{equation}    
The first term in the rhs is a combination of $\beta_{n}$ at 
$O(g_{n}^{L+1})$ and $\Gamma_{R(n)}^{(2,1)}$ 
at $O(g_{n}^{L})$, for $\beta_{n} \frac{\partial}{\partial g_{n}}$ 
is $O(g_{n})$. The two contributions coming from this term are finite at this 
loop order in the above equation. The last term has to do with 
$\Gamma_{R(n)}^{(2,2)}$ which has a skeleton expansion. Since the coupling 
constants and masses are actually finite at $O(g_{n}^{L})$, this implies that 
the last term is automatically finite at $O(g_{n}^{L+1})$. 
\par The second term in Eq.(46) has two 
contributions: $\Gamma_{R(n)}^{(2,1)}$ at $O(g_{n}^{L})$ which 
is hypothetically finite ($\gamma_{\phi^{2}(n)}$ starts at 
$O(g_{n})$) as well as $(\gamma_{\phi(n)}-\gamma_{\phi^{2}(n)})$ at 
$O(g_{n}^{L+1})$. The proof  that $(\gamma_{\phi(n)}-\gamma_{\phi^{2}(n)})$ 
is finite at $O(g_{n}^{L+1})$ still has to be demonstrated. The calculation 
of this expression at zero momentum is identical to Eq.(39) 
computed at $(L+1)$-loop order. The rhs of Eq.(39) 
involves the skeleton $\Gamma_{R(n)}^{(2,2)}$ at this loop order, which is 
finite. This shows that $(\gamma_{\phi(n)}-\gamma_{\phi^{2}(n)})$ is finite 
at $O(g_{n}^{L+1})$. Consequently, this concludes the proof that the rhs of 
Eq.(46) is finite.
\par Integration of the left hand side of Eq.(46) for arbitrary external 
momenta gives the result
\begin{equation}
(\mu_{n} \frac{\partial\Gamma_{R(n)}^{(2,1)}}{\partial \mu_{n}})|_{L+1} 
= f_{(n)}^{(2,1)}\; \;.
\end{equation}
Owing to the dimensionlessness of $\Gamma_{R(n)}^{(2,1)}$ at the critical 
dimension, we can perform a scaling transformation following similar steps  
to what was done for the vertex $\Gamma_{R(n)}^{(4)}$. By requiring
analyticity of $f_{(n)}^{(2,1)}$ for small $p_{n}$ and using the normalization 
conditions at zero external momenta, the proof that $\Gamma_{R(n)}^{(2,1)}$ 
is manifestly finite at $O(g_{n}^{L+1})$ follows at once. Using Eq.(37), we 
learn that $\gamma_{\phi(n)}$ is finite at $O(g_{n}^{L+1})$ as well.
\par The conclusion of the inductive proof of multiplicative 
renormalizability can be achieved by turning our attention to the case 
$N=2, M=0$ at order $g_{n}^{L+1}$ in (40), which reads
\begin{equation}
(\mu_{n} \frac{\partial\Gamma_{R(n)}^{(2)}}{\partial \mu_{n}})|_{L+1} = 
[-\beta_{n} \frac{\partial}{\partial g_{n}} 
\Gamma_{R(n)}^{(2)}]|_{L+1}
+ [\gamma_{\phi(n)}
\Gamma_{R(n)}^{(2,1)}]|_{L+1} + [\mu_{n}^{2n} (2n - \gamma_{\phi (n)}) \Gamma_{R(n)}^{(2,1)}]|_{L+1}   \;\;.
\end{equation}    
Now, $\Gamma_{R(n)}^{(2)}$ is required solely at $O(g_{n}^{L})$ 
(finite by hypothesis) since $\beta_{n}$ starts at $O(g_{n})$ and we have 
already proven that $\beta_{n}$ and $\gamma_{\phi(n)}$ are finite at 
$O(g_{n}^{L+1})$. This is sufficient to prove that the first two terms of the 
rhs in the above equation are manifestly finite. For the last term, we have 
just proved that both contributions inside it are finite at $O(g_{n}^{L+1})$. 
Since the rhs has a finite limit, we proceed to integrating the CSLE for this 
vertex function with arbitrary momenta. Perform the redefinition
\begin{equation}
\tilde{\Gamma}_{R(n)}^{(2)}(p_{n}) = 
\Gamma_{R(n)}^{(2)}(p_{n}) - p_{n}^{2n} - \mu_{n}^{2n}.
\end{equation}
The redefined vertex has mass dimension $\mu_{n}^{2n}$ and only 
deviates from $\Gamma_{R(n)}^{(2)}(p_{n})$ by higher powers of 
$p_{n}^{2 \tau}$ and $\mu_{n}^{2n}$, which turn out to be cancelled by 
negative powers of the cutoffs. Normalization conditions require that it 
vanishes with $(p_{n}^{2n})^{2}$ for small $|p_{n}|$.
Thus, collecting all the information contained in our previous 
discussions, $\tilde{\Gamma}_{R(n)}^{(2)}(p_{n})$ satisfies an 
equation of the type
\begin{equation}
[(\mu'_{n} \frac{\partial \tilde{\Gamma}_{R(n)}^{(2)}}{\partial \mu'_{n}})
(\frac{p_{n}}{\mu'_{n}}, \frac{\Lambda_{n}}{\mu'_{n}}, g_{n})]|_{L+1} = \mu_{n}^{' 2n}
f_{(n)}^{(2)}(\frac{p_{n}}{\mu'_{n}}, \frac{\Lambda_{n}}{\mu'_{n}}, g_{n}))|_{L+1} ,
\end{equation}
with $f_{(n)}^{(2)} = O((p_{n}^{4n}))$ for small $|p_{n}|$. 
Introduce the change of variables $\mu'_{n} = \frac{\mu_{n}}{\alpha}$. 
Integrate over the variable $\mu'_{n}$ in the interval 
$(\infty, \mu_{n})$. Using the normalization condition at zero external 
momenta as a boundary condition to the solution, i.e., 
$\tilde{\Gamma}_{R(n)}^{(2)}(0,0,g_{n}) = 0$, and taking the limit of 
infinite cutoffs, we get to
\begin{equation}
[\tilde{\Gamma}_{R(n)}^{(2)}(\frac{p_{n}}{\mu_{n}}, \infty, 
g_{n})]|_{L+1} = -\int_{0}^{1} \frac{d \alpha}{\alpha^{2n + 1}} 
[f_{(n)}^{(2)}(\alpha \frac{p_{n}}{\mu_{n}}, \infty, g_{n})]|_{L+1} .
\end{equation}
We are left with the task of proving that the integral is finite. But this is 
straightforward from the behavior of the integrand for small $p_{n}$: it goes 
like $f_{(n)}^{(2)} (\alpha \frac{p_{n}}{m_{n}}, \infty, g_{n})= 
O((\alpha p_{n})^{2n})^{2})$, which is free of singularities in the lower 
integration limit $\alpha \rightarrow 0$.
\par We have thus succeeded in demonstrating the multiplicative 
renormalizability by the inductive method as well as the existence of the 
Callan-Symanzik-Lifshitz Equations (36).  

\subsection{Isotropic behaviors}

The isotropic situation has a simple parallel with the $nth$ competition 
subspace of the anisotropic case as explicated before. We just have to keep 
in mind that the critical dimension is different $d_{c}=4n$ and there is only 
one type of competition along all space directions. Then it follows in a 
straightforward manner that the isotropic vertex parts satisfy exactly the 
Callan-Symanzik-Lifshitz equation associated to the $nth$ competing subspace 
of the anisotropic behaviors discussed above, namely 
\begin{eqnarray}
&&(\mu_{n} \frac{\partial}{\partial \mu_{n}} + 
\beta_{n} \frac{\partial}{\partial g_{n}}
- \frac{N}{2} \gamma_{\phi(n)} + L \gamma_{\phi^{2}(n)})
\Gamma_{R(n)}^{(N,L)} (p_{i (n)}, Q_{i (n)}, g_{n},
\mu_{n})= \\ \nonumber
&& \mu_{n}^{2n}(2n - \gamma_{\phi (n)}) \Gamma_{R(n)}^{(N,L+1)} (p_{i (n)}, 
Q_{i (n)};0, g_{n}, \mu_{n})  \;\;.
\end{eqnarray} 
\par In addition, the arguments presented above in the inductive proof of 
renormalizability for the $nth$ subspace in the anisotropic cases can also 
be used to prove inductively the finiteness of all vertex parts of the 
isotropic cases. The situation is identical to the proof furnished in the 
particular isotropic $m$-axial case $(n=2)$ \cite{CL}. By the same token, the 
existence of the Callan-Symanzik-Lifshitz equations (52) for arbitrary 
isotropic critical behaviors follows directly. This concludes the formal 
proof of multiplicative renormalizability of arbitrary competing systems of 
the Lifshitz type to all orders in perturbation theory.

\section{The Callan-Symanzik-Lifshitz equations at $d= d_{c} - \epsilon_{L}$}  

\par We will restrict our considerations only to vertex parts including 
composite operators which can be renormalized multiplicatively. The 
renormalized 1PI vertex parts are defined with respect to the bare functions 
in the following way 
 
\begin{eqnarray}
\Gamma_{R(n)}^{(N,L)} (p_{i (n)}, Q_{i(n)}, g_{n}, \mu_{n})
&=& Z_{\phi (n)}^{\frac{N}{2}} Z_{\phi^{2} (n)}^{L}
\Gamma^{(N,L)} (p_{i (n)}, Q_{i (n)}, \lambda_{0n}, \mu_{0n}, \Lambda_{n})\;.
\end{eqnarray}
\par We shall make reference to the cutoffs implicitly from now on. Although 
the regularization method utilizing cutoffs explained in the 
previous section is useful in many instances, we shall use dimensional 
regularization. Away from the critical dimension, the ultraviolet 
divergences of the original bare vertex parts can be represented by inverse 
powers (poles) in the variable $\epsilon_{L}$ in dimensional regularization. 
This manner of manipulating infinities will be shown in a moment to be 
particularly simple in the calculation of critical exponents.

\par The inductive proof of renormalizability of Eq.(53) at all orders in 
perturbation theory at the critical dimension (where the theory is 
renormalizable) was already demonstrated in the 
last section. Below the critical dimension, at $d=d_{c} - \epsilon_{L}$, the 
theory is less divergent (superrenormalizable) than its formulation at the 
critical dimension. Thus, we do not have to worry 
about the details of the rigorous proof of multiplicative renormalizability 
of the theory at $d=d_{c} - \epsilon_{L}$ in analogy with our discussion in 
the previous section at $d=d_{c}$, but take on this property as valid in that 
case as well. Instead, we shall give indirect evidence of this 
renormalizability by proving explicitly that the $\beta$-functions and Wilson 
functions are all finite at the specific perturbative order we are interested 
in the present work. 

\par It is interesting to express the renormalized and bare coupling constants 
in terms of the dimensionless couplings $u_{n}$, that is,
$g_{n} = u_{n} (\mu_{n}^{2n})^{\frac{\epsilon_{L}}{2}}$,
and $ \lambda_{0n} =  u_{0n} 
(\mu_{n}^{2n})^{\frac{\epsilon_{L}}{2}}$. Moreover, we can write all the 
renormalization functions in terms of $u_{n}$.

The dimensionless bare coupling constants $u_{0n}$ and the renormalization 
functions 
$Z_{\phi (n)}$, $\bar{Z}_{\phi^{2} (n)} = Z_{\phi (n)} Z_{\phi^{2} (n)}$ 
can be expanded in terms of the dimensionless parameter $u_{n}$ up 
to two-loop order as
\begin{subequations}
\begin{eqnarray}
&& u_{0n} = u_{n} (1 + a_{1n} u_{n} + a_{2n} u_{n}^{2}) ,\\
&& Z_{\phi (n)} = 1 + b_{2n} u_{n}^{2} + b_{3n} u_{n}^{3} ,\\
&& \bar{Z}_{\phi^{2} (n)} = 1 + c_{1n} u_{n} + c_{2n} u_{n}^{2} .
\end{eqnarray}
\end{subequations}
These ingredients can be used to discuss the Callan-Symanzik-Lifshitz 
equations analogously to the analysis performed at the critical dimension. 
They shall provide a better comprehension of the scaling behavior of the 
solutions in the anisotropic and isotropic sectors at the ultraviolet fixed 
points.

\subsection {Anisotropic}
\par There are at least two ways to derive the CSLE (at the and) away from the 
critical dimension 
$d = 4 + \sum_{n=2}^{L}[\frac{(n-1)}{n}]m_{n} - \epsilon_{L}$. In the first 
manner, one can take a total derivative of 
the renormalized vertex part (including composite operators) with respect to 
the logarithm of the renormalized mass $\mu_{n}$ characterizing the 
$m_{n}$-dimensional subspace at fixed bare coupling $\lambda_{n}$ and 
cutoff $\Lambda_{n}$, in conjunction with the normalization conditions. The 
second trend is to take a derivative of an arbitrary bare vertex part with 
relation to the bare  parameter $\mu_{0n}^{2n}$, and expressing everything 
solely in terms of renormalized quantities  via normalization conditions, 
with bare coupling $\lambda_{n}$ and 
cutoff $\Lambda_{n}$ kept fixed. The 
latter is almost identical to our procedure at the critical dimension 
$d_{c}= 4 + \sum_{n=2}^{L}[\frac{(n-1)}{n}]m_{n}$
explained above. Both strategies lead to the corresponding 
Callan-Symanzik-Lifshitz equations:
\begin{eqnarray}
&&(\mu_{n} \frac{\partial}{\partial \mu_{n}} + 
\beta_{n}(g_{n}, \mu_{n}) \frac{\partial}{\partial g_{n}}
- \frac{N}{2} \gamma_{\phi(n)} + L \gamma_{\phi^{2}(n)})
\Gamma_{R(n)}^{(N,L)} (p_{i (n)}, Q_{i (n)}, g_{n},
\mu_{n})= \\ \nonumber
&& \mu_{n}^{2n}(2n - \gamma_{\phi (n)}) \Gamma_{R(n)}^{(N,L+1)} (p_{i (n)}, 
Q_{i (n)};0, g_{n}, \mu_{n})  \;\;.
\end{eqnarray} 
\par These equations resemble the CSLE at the critical dimension, but now 
the renormalized coupling constants $g_{n}$ are dimensionful. This provokes 
a discrepancy between the beta functions 
defined at the critical dimension and its corresponding version at 
$\epsilon_{L} \neq 0$. When expressed in terms of the renormalized parameters, 
the coefficients appearing in the CSLE equations satisfy the following 
relations:
\begin{equation}
\beta_{n} (g_{n}, \mu_{n})= \mu_{n} \frac{\partial g_{n}}{\partial \mu_{n}}, 
\end{equation}
\begin{equation}
\gamma_{\phi(n)}(g_{n}, \mu_{n}) = \mu_{n} \frac{\partial ln Z_{\phi (n)}}{\partial \mu_{n}},
\end{equation}
\begin{equation}
\gamma_{\phi^{2}(n)}(g_{n}, \mu_{n}) = - \mu_{n} \frac{\partial ln Z_{\phi^{2} (n)}}{\partial \mu_{n}}. 
\end{equation}
\par Replacing the definition 
$g_{n} = u_{n} (\mu_{n}^{2n})^{\frac{\epsilon_{L}}{2}}$ into the above 
equation for $\beta_{n}(g_{n}, \mu_{n})$ we find 
\begin{equation}
\beta_{n}(g_{n}, \mu_{n}) \frac{\partial}{\partial g_{n}} = 
(\mu_{n} \frac{\partial u_{n}}{\partial \mu_{n}})_{\lambda_{0n}} 
\frac{\partial}{\partial u_{n}} + 
n \epsilon_{L} g_{n} \frac{\partial}{\partial g_{n}} \;\; ,
\end{equation}
where the derivative in the first term is calculated at fixed bare coupling 
constant $\lambda_{0n}$. It is equivalent to the 
dimensionless function $\beta_{n}(u_{n}) = (\mu_{n} \frac{\partial u_{n}}{\partial \mu_{n}})_{\lambda_{0n}}$. Now define the Gell-Mann-Low 
function for $\epsilon_{L} \neq 0$ in terms of the dimensionless 
coupling according to \cite{Vladimirov,Naud}  
\begin{equation}
[\beta_{n}(g_{n}, \mu_{n})]_{GL}= -n \epsilon_{L} g_{n} 
+ \beta_{n}(g_{n}, \mu_{n}) \;\;,
\end{equation}
where the last term can be identified with the value of the function at the 
critical dimension. The solution of the Callan-Symanzik-Lifshitz equations 
for the vertex parts  away from the critical dimension will possess a scaling 
limit provided it can be expressed entirely in terms of dimensionless 
quantities. We can get rid of the undesirable contribution coming from 
dimensionful couplings if we start from scratch with the Gell-Mann-Low 
function in Eq.(64). In that case, Eq.(59) turns out to be   
\begin{equation}
[\beta_{n}]_{GL}(g_{n}) \frac{\partial }{\partial g_{n}} = 
\beta_{n}(u_{n}) \frac{\partial }{\partial u_{n}}.
\end{equation}   
\par Consequently, the Callan-Symanzik-Lifshitz equations away from the critical dimension 
can be rewritten as
\begin{eqnarray}
&&(\mu_{n} \frac{\partial}{\partial \mu_{n}} + 
\beta_{n}(u_{n}) \frac{\partial}{\partial u_{n}}
- \frac{N}{2} \gamma_{\phi(n)} + L \gamma_{\phi^{2}(n)})
\Gamma_{R(n)}^{(N,L)} (p_{i (n)}, Q_{i (n)}, u_{n},
\mu_{n})= \\ \nonumber
&& \mu_{n}^{2n}(2n - \gamma_{\phi (n)}) \Gamma_{R(n)}^{(N,L+1)} (p_{i (n)}, 
Q_{i (n)};0, u_{n}, \mu_{n})  \;\;.
\end{eqnarray}  
\par Let us take a closer look at the dimensionless functions 
$\beta_{n}(u_{n})$. Since the derivatives of the renormalized masses 
in terms of dimensionless renormalized coupling constants  are taken 
at fixed bare coupling constant, we can use the identity
\begin{equation}
\mu_{n} \frac{\partial u_{n}}{\partial \mu_{n}}= - \mu_{n} (\frac{\partial \lambda_{0n}}{\partial \mu_{n}})_{u_{n}} \;(\frac{\partial u_{n}}{\partial \lambda_{0n}})_{\mu_{n}}\;\;,
\end{equation} 
in order to trade partial derivatives. Using last equation, we find the 
following result for the flow functions in terms of dimensionless 
parameters    
\begin{equation}
\beta_{n}(u_{n}) = - n \epsilon_{L}(\frac{\partial ln u_{0n}}{\partial u_{n}})^{-1}.
\end{equation}
Employing this result, the remaining renormalization functions can be 
rewritten in terms of dimensionless quantities in the form
\begin{subequations}
\begin{eqnarray}
&& \gamma_{\phi (n)}(u_{n})  = \beta_{n}
\frac{\partial ln Z_{\phi (n)}}{\partial u_{n}}\\
&& \gamma_{\phi^{2} (n)}(u_{n}) = - \beta_{n}
\frac{\partial ln Z_{\phi^{2} (n)}}{\partial u_{n}}\\
&& \bar{\gamma}_{\phi^{2} (n)}(u_{n}) = - \beta_{n}
\frac{\partial ln \bar{Z}_{\phi^{2} (n)}}{\partial u_{n}}  .
\end{eqnarray}
\end{subequations}
\par The equations (62) are the Callan-Symanzik-Lifshitz equations for 
anisotropic vertex parts describing generic competing systems of Lifshitz 
type with arbitrary composite operators which are multiplicatively 
renormalizable. Note that $\Gamma_{R(n)}^{(0,2)}$ is precluded from this 
discussion. The independence of renormalization flows in each subspace 
characterized by their corresponding bare(/renormalized) masses and coupling 
constants are manifest in this construction of anisotropic criticalities. 

The main obstacle to encounter a general solution of the CSL Eq.(62) is 
the appearance of the inhomogeneous term $\Gamma_{R(n)}^{(N,L+1)} (p_{i (n)}, 
Q_{i (n)},0, g_{n}, \mu_{n})$ in its right hand side (rhs). It corresponds to 
the insertion at zero momentum in the vertex 
$\Gamma_{R(n)}^{(N,L)} (p_{i (n)}, Q_{i (n)}, g_{n},\mu_{n})$ due to the 
action of the derivative with respect to the mass parameter characterizing 
the $n$-th competing subspace in the anisotropic cases. For the sake of 
simplicity, take $L=0$. Generally speaking, this derivative produces an 
additional propagator in $\Gamma_{R(n)}^{(N,L+1)} (p_{i (n)}, 
Q_{i (n)},0, g_{n}, \mu_{n})$ when we compare it with
$\Gamma_{R(n)}^{(N,L)} (p_{i (n)}, Q_{i (n)}, g_{n},\mu_{n})$. The limit 
$p_{i(n)} \rightarrow \infty$ describes the ultraviolet behavior of such 
vertex parts.  Thus, $\Gamma_{R(n)}^{(N,1)} (p_{i (n)},0, g_{n}, \mu_{n})$ 
is of order $p_{i (n)}^{-2n}\Gamma_{R(n)}^{(N)} (p_{i (n)}, g_{n}, \mu_{n})$ 
for large $p_{n}$, up to powers of $lnp_{n}$, at every order in  
perturbation theory. We make the assumption that these logarithms do not 
add up to compensate the factor $p_{i (n)}^{-2n}$. 

All the momenta involved in a given Feynman diagram are Euclidean. Taking a 
fixed nonvanishing  Euclidean momentum $k_{i(n)}$, we can reach the 
ultraviolet region of any diagram by performing a scale transformation 
like $p_{i(n)}= \rho_{n} k_{i(n)}$, where $\rho_{n}$ is a dimensionless flow 
variable in the limit $\rho_{n} \rightarrow \infty$. Interesting anisotropic 
configurations have the following structure: there are $n$ independent 
subsets of momenta which go to infinity independently. Since the $n$ 
independent mass subspaces have a well defined zero mass limit \cite {generic2} 
and all the momenta appearing in 
$\Gamma_{R(n)}^{(N)}(p_{i (n)}, g_{n}, \mu_{n})$ 
are nonexceptional in the ultraviolet regime (except for the zero momentum 
of the inserted $\phi^{2}$ operator in the vertex part 
$\Gamma_{R(n)}^{(N,1)}(p_{i (n)},0, g_{n}, \mu_{n})$), we can apply 
the Weinberg's theorem \cite{Weinberg} in each subspace, 
which guarantees that the rhs of Eq.(62) can be neglected order by 
order in the perturbative expansion. 

The limit $\rho_{n} \rightarrow \infty$ implies to take all internal 
momenta in a given diagram as large as possible. This goal can be scored by 
regulating the integral with cutoffs $\Lambda_{n}$ in the limit 
$\Lambda_{n} \rightarrow \infty$. Hence, the integral becomes a 
homogeneous function of the mass $\mu_{n}$. 
The regions in momentum space where the rhs in Eq. (62) can be neglected when 
compared to the left-hand side (lhs) come from the limit 
$\frac{p_{i (n)}}{\mu_{n}} \rightarrow \infty$ which is identical to the 
ultraviolet limit. This is indeed the case, as we have already shown in 
Sec. III using cutoffs. Moreover, it will be demonstrated in the Appendixes 
via dimensional regularization that the ultraviolet divergences are associated 
to this region in momenta space where the loop integrals shall be computed.

Recall that several dimensional redefinitions of the momentum components 
along the various competing subspaces have been performed as discussed 
in Sec. II (see also \cite{generic2}). If $d$ is the spatial dimension of the 
system, these redefinitions along with the Lifshitz conditions translate 
themselves into an ``effective'' space dimension for the anisotropic 
situations of generic competing systems, i.e., 
$d_{eff}= d - \sum_{n=2}^{L}[\frac{n-1}{n}]m_{n}$. Elementary dimensional 
analysis implies that, under scaling, the vertex parts possess the following 
behavior

\begin{eqnarray}
\Gamma_{R (\tau)}^{(N)} (\rho_{n} k_{i (n)}, u_{n}, \mu_{n})&=&
\rho_{n}^{n(N + (d - \sum_{n=2}^{L}[\frac{n-1}{n}]m_{n}) 
- \frac{N(d - \sum_{n=2}^{L}[\frac{n-1}{n}]m_{n})}{2})}
\;\;\Gamma_{R (n)}^{(N)} (k_{i (n)}, u_{n}(\rho_{n}),
\frac{\mu_{n}}{\rho_{n}})\nonumber.
\end{eqnarray}

The solution of the asymptotic part associated to the vertex function 
satisfying the homogeneous CSL equations has the property 
\begin{eqnarray}
\Gamma_{as\; R (n)}^{(N)} (k_{i (n)}, u_{n}, \frac{\mu_{n}}{\rho_{n}})&=&
exp[-\frac{N}{2} \int_{u_{n}}^{u_{n}(\rho_{n})} \gamma_{\phi (n)}(u'_{n}(\rho_{n}))
\frac{{d u'_{n}}}{\beta_{n}(u'_{n})}]\\
&&\Gamma_{as\; R (n)}^{(N)} (k_{i (n)}, u_{n}(\rho_{n}), 
\mu_{n})\nonumber, 
\end{eqnarray}
where 
\begin{equation}
\rho_{n}= \int_{u_{n}}^{u_{n}(\rho_{n})} \frac{du'_{n}}{\beta_{n}(u'_{n})} \; \; .
\end{equation}
From now on we omit the subscript for asymptotic in the vertex parts 
satisfying the homogeneous Callan-Symanzik-Lifshitz equations. The eigenvalue 
conditions $\beta_{n}(u_{n \infty})=0$ yields the ultraviolet nontrivial 
fixed point. Exactly at this value of the coupling, which is independent 
of the competing subspace under consideration in the anisotropic cases, 
the solutions of the CSL Eq.(62) under a scale in the external momenta 
can be written as
\begin{eqnarray}
\Gamma_{R (n)}^{(N)} (\rho_{n} k_{i (n)}, u_{n \infty}, \mu_{n})&=&\rho_{n}^{n (N + (d - \sum_{n=2}^{L}[\frac{n-1}{n}]m_{n}) 
- \frac{N(d - \sum_{n=2}^{L}[\frac{n-1}{n}]m_{n})}{2})
-\frac{N \gamma_{\phi (n)}(u_{n \infty})}{2} } \\
&&\Gamma_{R (n)}^{(N)} (k_{i (n)}, u_{n \infty}, \mu_{n})\nonumber .
\end{eqnarray}
The dimension of the field can be defined by
\begin{equation}
\Gamma_{(n)}^{(N)}(\rho_{n}k_{i (n)}) = 
\rho_{n}^{n [(d - \sum_{n=2}^{L}[\frac{n-1}{n}]m_{n}) - N d_{\phi (n)}]}
\Gamma_{(n)}^{(N)}(k_{i}).
\end{equation}
Due to the interactions, the field develop an anomalous term, which can also 
be defined through the relation 
$d_{\phi (n)}= \frac{d - \sum_{n=2}^{L}[\frac{n-1}{n}]m_{n}}{2} - 1 + 
\frac{\eta_{n}}{2n}$. Comparing Eqs. (68) and (69), one can easily verify 
that the anomalous dimension $\eta_{n}$  can be identified with 
$\gamma_{\phi (n)}(u_{n \infty})$, i.e., 
$\eta_{n}=\gamma_{\phi (n)}(u_{n \infty})$. 

Plus, the anomalous dimension of the composite operator $\phi^{2}$ can be 
extracted by analyzing the vertex parts including composite operators. The 
asymptotic behavior of the vertex parts at the fixed point 
$u_{n}=u_{n \infty}$ coming from the solution of the CSL Eq.(62) has the 
following simple scaling property $((N,L) \neq (0,2))$
\begin{eqnarray}
\Gamma_{R (n)}^{(N,L)} (\rho_{n} k_{i (n)},\rho_{n} p_{i (n)} , u_{n \infty}, 
\mu_{n})&=&\rho_{n}^{n(N + (d - \sum_{n=2}^{L}[\frac{n-1}{n}]m_{n}) 
- \frac{N(d - \sum_{n=2}^{L}[\frac{n-1}{n}]m_{n})}{2} -2L)-\frac{N\gamma_{\phi (n)}(u_{n \infty})}{2}}\nonumber\\
&&\times\;\; \rho_{n}^{L\gamma_{\phi^{2} (n)}(u_{n \infty})} 
\Gamma_{R (n)}^{(N)} (k_{i (n)}, p_{i (n)} , u_{n \infty}, \mu_{n}).
\end{eqnarray}
If we write the coefficient in the rhs as 
$\rho_{n}^{n[(d - \sum_{n=2}^{L}[\frac{n-1}{n}]m_{n}) - N d_{\phi (n)}] + L d_{\phi^{2} (n)}}$, 
we discover that 
$d_{\phi^{2} (n)} = -2n + \gamma_{\phi^{2} (n)}(u_{n \infty})$. The 
correlation length exponents can be computed through the identification 
$\nu_{n}^{-1}= - d_{\phi^{2} (n)}= 2n - \gamma_{\phi^{2} (n)}(u_{n \infty})$. 

This method with several independent mass scales generalizes, in a nontrivial 
way, the previous method developed for $m$-axial anisotropic Lifshitz points 
where only two mass scales are present.    

\subsection{Isotropic}

The existence of only one type of competing axis in the isotropic 
criticalities make the analysis simpler for the cases $d=m_{n}$. As before, 
we shall briefly discuss this case by focusing on the analogy with the 
competing sector labeled by $n$ of the anisotropic treatment already discussed 
above. We start with the mass scale $\mu_{n}$, dimensionless coupling 
constant $u_{n}$ and subscript $n$ in all vertex functions. The critical 
dimension is $4n$, whereas the expansion parameter is 
$\epsilon_{L}= 4n - d$. The effective space dimension is 
$\frac{m_{n}}{n}$. At the (eigenvalue condition $\beta_{n}(u_{n \infty})=0$/) 
fixed point $u_{n \infty}$, a scale transformation in the external momenta 
implies that the vertex functions present the following property:
\begin{eqnarray}
\Gamma_{R (n)}^{(N)} (\rho_{n} k_{i (n)}, u_{n \infty}, \mu_{n})&=&\rho_{n}^{n(N 
+ \frac{m_{n}}{n} - N \frac{m_{n}}{2n})
-\frac{N \gamma_{\phi (n)}(u_{n \infty})}{2} } \\
&&\Gamma_{R (n)}^{(N)} (k_{i (3)}, u_{n \infty},m_{3})\nonumber .
\end{eqnarray}
The dimension of the field is defined by
\begin{equation}
\Gamma_{(n)}^{(N)}(\rho_{n}k_{i (n)}) = \rho_{n}^{n[(\frac{m_{n}}{n}) - N d_{\phi (n)}]}
\Gamma_{(n)}^{(N)}(k_{i}).
\end{equation}
Then, it follows that 
$d_{\phi (n)}= \frac{m_{n}}{2n} - 1 + \frac{\eta_{n}}{2n}$, which in turn 
implies that the anomalous dimension of the field  
($\eta_{n} \equiv \eta_{Ln}$) satisfies the identity 
$\eta_{n}=\gamma_{\phi (n)}(u_{n \infty})$. 
\par Now, let us consider the vertex parts including composite fields. The 
identification of the anomalous dimension of the composite 
operator $\phi^{2}$ can be done through the following steps: at the fixed 
point $u_{n}=u_{n \infty}$ the asymptotic behavior of the vertex parts under 
scale transformation can be written as 
\begin{eqnarray}
\Gamma_{R (n)}^{(N,L)} (\rho_{n} k_{i (n)},\rho_{n} p_{i (n)} , u_{n \infty}, 
\mu_{n})&=&
\rho_{n}^{n(N + \frac{m_{n}}{n} - N\frac{m_{n}}{2n}) -2L)
-\frac{N\gamma_{\phi (n)}(u_{n \infty})}{2}+ L\gamma_{\phi^{2} (n)}(u_{n \infty})}\\ 
&&\times\;\;\Gamma_{R (n)}^{(N)} (k_{i (n)}, p_{i (n)} , u_{n \infty},m_{n})\nonumber    .
\end{eqnarray}
Then, write the coefficient in the rhs as 
$\rho_{n}^{n[\frac{m_{n}}{n} - N d_{\phi (n)}] + L d_{\phi^{2} (n)}}$. One 
learns that $d_{\phi^{2} (n)} = -2n + \gamma_{\phi^{2} (n)}(u_{n \infty})$ and 
the correlation length exponent $(\nu_{n} \equiv \nu_{Ln})$ in the isotropic 
case can be identified with $d_{\phi^{2} (n)}$ through  
$\nu_{n}^{-1}= - d_{\phi^{2} (n)}= 2n - \gamma_{\phi^{2} (n)}(u_{n \infty})$.
\par Finally, the beta function does not have the 
global factor of $n$ as in the competition directions of the anisotropic case. 
Rather, in terms of dimensionless parameters $\beta_{n} = -  \epsilon_{L}(\frac{\partial ln u_{0 n}}{\partial u_{n}})^{-1}$. 
\par These are all informations required for the calculations of the critical 
exponents by diagrammatic means. This theme will be developed in the following 
remaining sections. 

\section{Critical exponents in the anisotropic cases}

In order to calculate the critical exponents for anisotropic critical 
behaviors of arbitrary competing systems, we shall use the results in 
Appendix A for the massive Feynman graphs using the generalized orthogonal 
approximation. As already demonstrated in \cite{generic1, generic2}, this 
approximation 
is the most general one consistent with homogeneity in the several 
independent scales of external momenta. In the massive case introduced in 
the present work, the orthogonal approximation is also consistent with 
homogeneity in the various mass scales associated to each competing 
subspace. We shall employ the normalization conditions setup of Sec.II 
in our computational procedure.

To begin with, we use Eqs.(3) and (53) to fixing the normalization functions
$Z_{\phi (n)}, \bar{Z}_{\phi^{2} (n)}$ in powers of $u_{n}$ in conjumination 
with the appropriate Feynman diagrams at the loop order needed. In the 
expressions of the one- two and three-loop level of the integrals 
$I_{2}, I'_{3}, I_{4}, I'_{5}$ given in Appendix A, we emphasize that a 
geometric angular factor appears every time we perform a loop integral. It 
is explicitly given by $[S_{(d-\sum_{n=2}^{L}m_{n})} 
\Gamma(2 - \sum_{n=2}^{L}\frac{m_{n}}{2n})(\Pi_{n=2}^{L} 
\frac{S_{m_{n}} \Gamma(\frac{m_{n}}{2n})}{2n})]$ and can be absorbed in a 
redefinition of the coupling constants. We can perform those redefinitions 
in order to discard it from our considerations henceforth. We then 
obtain the renormalization functions in terms of those loop integrals in 
the following form 
\begin{subequations}
\begin{eqnarray}
&& u_{0 n}= u_{n}[1 + \frac{(N+8)}{6} I_{2} u_{n} + 
(\frac{[(N+8)I_{2}]^{2}}{18}\nonumber \\ 
&& - (\frac{(N^{2}+6N+20)I_{2}^{2}}{36} + 
\frac{(5N+22)I_{4}}{9}) - \frac{(N+2)I'_{3}}{9})u_{n}^{2}],\\
&& Z_{\phi (n)}= 1 + \frac{(N+2)I'_{3}}{18} u_{n} 
+ \frac{(N+2)(N+8)(I_{2}I'_{3} - \frac{I'_{5}}{2})}{54}u_{n}^{2},\\
&& \bar{Z}_{\phi^{2} (n)}= 1 + \frac{(N+2)I_{2}}{6} u_{n}\nonumber \\
&& + [\frac{(N^{2}+7N+10)I_{2}}{18} 
- \frac{(N+2)}{6}(\frac{(N+2)I_{2}^{2}}{6} 
+ I_{4})]u_{n}^{2}.
\end{eqnarray}
\end{subequations}
In dimensional regularization, the coefficients of the various terms in 
powers of $u_{n}$ present poles in $\epsilon_{L}$. They cancel in the 
evaluation of  $\beta_{n}$ and the critical exponents. When we combine 
Eqs.(54), (64) and (65) together, we obtain the 
$\beta_{n}$ and Wilson functions in every subspace. We can write them through 
the following expressions
\begin{subequations}
\begin{eqnarray}
&& \beta_{n}  =  -n \epsilon_{L}u_{n}[1 - a_{1 n} u_{n}
+2(a_{1 n}^{2} -a_{2 n}) u_{n}^{2}],\\
&& \gamma_{\phi (n)} = -n \epsilon_{L}u_{n}[2b_{2 n} u_{n}
+ (3 b_{3 n}  - 2 b_{2 n} a_{1 n}) u_{n}^{2}],\\
&& \bar{\gamma}_{\phi^{2} (n)} = n \epsilon_{L}u_{n}[c_{1 n}
+ (2 c_{2 n}  - c_{1 n}^{2} - a_{1 n} c_{1 n})u_{n}    ].
\end{eqnarray}
\end{subequations}
Using the explicit values of the integrals presented in Appendix A, we can 
determine the above mentioned coefficients by using the normalization 
conditions as functions of those integrals calculated at zero external 
momenta. When the results in Appendix A are combined with Eqs.(54) and (74), 
we conclude therefore that
\begin{subequations}
\begin{eqnarray}
&& a_{1 n} = \frac{(N+8)}{6 \epsilon_{L}}[1 + (h_{m_{L}}-1) \epsilon_{L}] ,\\
&& a_{2 n} = (\frac{N+8}{6 \epsilon_{L}})^{2}
+ [\frac{(N+8)^{2}}{18}(h_{m_{L}} - 1) - \frac{(3N+14)}{24}]
\frac{1}{\epsilon_{L}} ,\\
&& b_{2 n} = -\frac{(N+2)}{144 \epsilon_{L}}[1 +
(2 h_{m_{L}} - \frac{5}{4}) \epsilon_{L}] - \frac{(N+2)}{144}I , \\
&& b_{3 n} = -\frac{(N+2)(N+8)}{1296 \epsilon_{L}^{2}} +
\frac{(N+2)(N+8)}{108 \epsilon_{L}}(-\frac{1}{4} h_{m_{L}} +
\frac{13}{48}), \\
&& c_{1 n} = \frac{(N+2)}{6 \epsilon_{L}}[1 + (h_{m_{L}} - 1) \epsilon_{L}], \\
&& c_{2 n} = \frac{(N+2)(N+5)}{36 \epsilon_{L}^{2}}
+ \frac{(N+2)(N+5)}{18 \epsilon_{L}}(h_{m_{L}} - 2) + \frac{(4N^{2} + 25N +34)}{72 \epsilon_{L}}.
\end{eqnarray}
\end{subequations}
Just as in the case of noncompeting and $m$-axial universality classes, the 
integral I encountered in Appendix Eq.(A9) appears explicitly in the 
coefficient $b_{2 n}$ Eq. (76c). We emphasize that these expressions differ 
from Eqs.(67) of Ref.\cite{generic2} obtained in the massless theory at 
nonvanishing external momenta in each subspace. Since the normalization 
conditions in the present work are defined at zero external momenta as well 
as nonvanishing mass in the specific subspace labeled by $n$ and recalling 
that these coefficients depend upon the normalization conditions, this 
disagreement in their values using either renormalization scheme should not be 
surprising.  

Inserting the values of the coefficients given in Eqs. (76) into (75a), 
$\beta_{n}$ can be written as 
\begin{eqnarray}
\beta_{n}=&& -n u_{n}[\epsilon_{L} 
- \frac{(N+8)}{6}(1+(h_{m_{L}} -1))u_{n} \nonumber\\
&&  - \frac{(3N+14)}{12}u_{n}^{2}] + O(u_{n}^{4}).
\end{eqnarray}
Using last equation, the eigenvalue conditions $\beta_{n}(u_{n \infty})=0$ 
lead to two solutions: a trivial zero as well as a nontrivial zero of order 
$\epsilon_{L}$ of each 
$\beta_{n}$ characterizing the independent noncompeting ($n=1$) and 
competing ($n=2,...,L$) subspaces. Remarkably, they correspond to the same 
value of the coupling constant 
($u_{1 \infty}=u_{2 \infty}=...=u_{L \infty} \equiv u_{n \infty})$, i.e.,
\begin{equation}
u_{n \infty}=\frac{6}{8 + N}\,\epsilon_L\Biggl\{1 + \epsilon_L
\,\Biggl[ - (h_{m_{L}} -1) + \frac{(9N + 42)}{(8 + N)^{2}}\Biggr]\Biggr\}\;\;.
\end{equation}   
Substitution of this value in the functions $\gamma_{\phi (n)}$ and 
$\bar{\gamma}_{\phi^{2} (n)}$ together with the coefficients given in 
Eqs.(76) allows us to find the critical exponents $\eta_{n}$ and 
$\nu_{n}$. Indeed procceding as indicated, we first find
\begin{eqnarray}
&&\gamma_{\phi(n)}(u_{n \infty})= \frac{n}{2} \epsilon_{L}^{2}\,\frac{N + 2}{(N+8)^2}
[1 + \epsilon_{L}(\frac{6(3N + 14)}{(N + 8)^{2}} - \frac{1}{4})] .
\end{eqnarray}
Note that these expressions are exactly the same of those obtained 
previously in \cite{generic2}, Eq.(73) therein, upon the identification 
$\gamma_{\phi(n)}(u_{n \infty})=\eta_{n}$.

The exponents $\nu_{n}$ can be encountered from the expression of the 
anomalous dimension of the composite operator $\phi^{2}$. As we know, 
$\gamma_{\phi^{2}(n)}= \bar{\gamma}_{\phi^{2}(n)}+ \gamma_{\phi (n)}$. Thus, 
it follows that $\nu_{n}^{-1}= - d_{\phi^{2} (n)}= 2 n 
- \gamma_{\phi^{2} (n)}(u_{n \infty}) 
- \gamma_{\phi(n)}(u_{n \infty})$. Using Eqs.(76) once more, the following 
intermediate result can be verified:
\begin{equation}
\bar{\gamma}_{\phi^{2} (n)}(u_{n}) = n \frac{(N+2)}{6} u_{n}
[1 + \epsilon_{L}(h_{m_{L}} - 1) - \frac{1}{2} u_{n}].
\end{equation}
Using the value $u_{n \infty}$ into the last expression along with 
$\gamma_{\phi(n)}(u_{n \infty})$ from Eq.(79), we get 
\begin{eqnarray}
&& \nu_{n} =\frac{1}{2n} + \frac{(N + 2)}{4 n (N + 8)} \epsilon_{L}
+  \frac{1}{8 n}\frac{(N + 2)(N^{2} + 23N + 60)} {(N + 8)^3} \epsilon_{L}^{2}.
\end{eqnarray}
These exponents are in exact agreement with those calculated 
previously using a massless theory \cite{generic2}. Note that the properties 
of strong anisotropic scaling \cite{Henkel} 
$\eta_{n}= n \eta_{1}$ and 
$\nu_{n}= (\frac{1}{n}) \nu_{1}$ already encountered in the massless case are 
reproduced using quite a different method and generalizes the more restricted 
$m$-axial Lifshitz universality class \cite{Picture,CL}.

\section{Isotropic critical exponents in the orthogonal approximation}

Although the isotropic diagrams can be calculated exactly without the need of 
any sort of approximation, in this section we shall evaluate the isotropic 
exponents for two reasons. Firstly, for completeness: pursuing the analogy 
with the anisotropic cases within an approximation which keeps the desirable 
concept of homogeneity in the mass(es). Secondly, simplicity is a key 
ingredient, since possible applications in other field theories with higher 
derivatives might be worthwhile using this simple setting of computation. 
Due to nontrivial features of the exact computation, we postpone the 
discussion of the calculation of the integrals and of the exact critical 
exponents to the next sections.  

We now have only one type of subspace to be integrated over, along with only 
one type of mass parameter for each isotropic behavior labeled by the number 
of neighbors $n$ coupled via competing interactions according to the CECI 
model introduced in \cite{generic1} to understanding the most general 
magnetic systems with competing interactions. We can take advantage of the 
discussion 
in the previous section: we focus on the $nth$ subspace taking into account 
of course the particularities of the isotropic renormalization functions as 
pointed out in Sec. V. In addition, we shall use the results in Appendix B 
for the Feynman integrals in order to compute the exponents $\eta_{n}$ and 
$\nu_{n}$ via perturbative expansion. Each loop integral originates an 
angular factor of $S_{m_{n}}$, the area of an $m_{n}$-dimensional unit sphere, 
which shall be absorbed in a redefinition of the coupling constant. In order 
to make a clear distinction among the various isotropic behaviors and the 
$nth$ competing subspace of the anisotropic behaviors, we shall replace 
the perturbative parameter in the isotropic case, i.e., 
$\epsilon_{L} \rightarrow  \epsilon_{n}$. Furthermore, this slight change 
of notation shall be useful to perform a comparison with the critical 
exponents already evaluated in the massless theory in Ref.\cite{generic2}.  

First we list the analogues of Eqs.(75) for the renormalization functions in 
the isotropic cases, namely 
\begin{subequations}
\begin{eqnarray}
&& \beta_{n}  =  -\epsilon_{n}u_{n}[1 - a_{1n} u_{n}
+2(a_{1n}^{2} -a_{2n}) u_{n}^{2}],\\
&& \gamma_{\phi (n)} = - \epsilon_{n}u_{n}[2b_{2n} u_{n}
+ (3 b_{3n}  - 2 b_{2n} a_{1n}) u_{n}^{2}],\\
&& \bar{\gamma}_{\phi^{2} (n)} = \epsilon_{n}u_{n}[c_{1n}
+ (2 c_{2n}  - c_{1n}^{2} - a_{1n} c_{1n})u_{n}    ].
\end{eqnarray}
\end{subequations} 
Employing the results of the previous section along with the output 
obtained in Appendix B utilizing the orthogonal approximation in the 
computation of the appropriate loop integrals for the isotropic behaviors, 
the various coefficients can be expressed in the form
\begin{subequations}
\begin{eqnarray}
&& a_{1 n} = \frac{N+8}{6 \epsilon_{n}}[1 - \frac{1}{2n} \epsilon_{n}] ,\\
&& a_{2 n} = (\frac{N+8}{6 \epsilon_{n}})^{2}
- [\frac{2N^{2} + 41N + 170}{72n \epsilon_{n}}] ,\\
&& b_{2 n} = -\frac{(N+2)}{144n \epsilon_{n}}[1 - \frac{1}{4n} \epsilon_{n}] 
- \frac{(N+2)}{144 n^{2}} I, \\
&& b_{3 n} = -\frac{(N+2)(N+8)}{1296n \epsilon_{n}^{2}} +
\frac{7(N+2)(N+8)}{5184 n^{2} \epsilon_{n}}, \\
&& c_{1 n} = \frac{(N+2)}{6 \epsilon_{n}}[1 - \frac{1}{2n} \epsilon_{n}], \\
&& c_{2 n} = \frac{(N+2)(N+5)}{36 \epsilon_{n}^{2}}
- \frac{(N+2)(2N+13)}{72 n \epsilon_{n}}.
\end{eqnarray}
\end{subequations}
Note that these expressions are different from Eqs.(90) 
derived in Ref.\cite{generic2} using the orthogonal 
approximation in the massless limit. Since $\epsilon_{n}=4n -d$ and 
substituting the above results in the expression for $\beta_{n}$, we find  
\begin{eqnarray}
\beta_{n}=&& - u_{n}[\epsilon_{n} 
- \frac{(N+8)}{6}(1 - \frac{1}{2n} \epsilon_{n})u_{n} \nonumber\\
&&  + \frac{(3N+14)}{12n}u_{n}^{2}] + O(u_{n}^{4}).
\end{eqnarray}
The eigenvalue condition $\beta_{n}(u_{n \infty})=0$ permits us to find out 
the nontrivial fixed point of the dimensionless coupling constant, whose 
value is given by
\begin{equation}
u_{n \infty}=\frac{6}{8 + N}\,\epsilon_{n}\Biggl\{1 + \frac{\epsilon_{n}}{n}
\,\Biggl[ \frac{1}{2} + \frac{(9N + 42)}{(8 + N)^{2}}\Biggr]\Biggr\}\;\;.
\end{equation}
Replacing this fixed point in the expression for $\gamma_{\phi (n)}$ together 
with Eqs.(83), we then have
\begin{eqnarray}
&&\gamma_{\phi(n)}(u_{n \infty})= \epsilon_{n}^{2}\,\frac{N + 2}{2 n (N+8)^2}
[1 + \epsilon_{n} \frac{1}{n}(\frac{6(3N + 14)}{(N + 8)^{2}} - \frac{1}{4})] .
\end{eqnarray}
This value corresponds exactly to the exponent $\eta_{n}$ previously 
obtained using the renormalization group equation in the massless 
theory \cite{generic1,generic2}. Furthermore, using again the results (83)  
in the definition of $\bar{\gamma}_{\phi^{2} (n)}$, we obtain
\begin{equation}
\bar{\gamma}_{\phi^{2} (n)}(u_{n}) = \frac{(N+2)}{6} u_{n}
[1 -\frac{1}{2 n} \epsilon_{n} - \frac{1}{2 n} u_{n}].
\end{equation}
Replacing  Eq.(85) into last equation, we get to the following result
\begin{equation}
\bar{\gamma}_{\phi^{2} (n)}(u_{n \infty}) = \frac{(N+2)}{(N+8)} \epsilon_{n}
[1 +\frac{6(N+3)}{n (N+8)^{2}}\epsilon_{n} ].
\end{equation}
Using the relation $\nu_{n}=(2n-\bar{\gamma}_{\phi^{2} (n)}(u_{n \infty})-\gamma_{\phi(n)}(u_{n \infty}))^{-1}$, we find out the correlation length critical 
exponent 
\begin{eqnarray}
&& \nu_{n} =\frac{1}{2n} + \frac{(N + 2)}{4 n^{2} (N + 8)} \epsilon_{n}
+  \frac{1}{8 n^{3}}\frac{(N + 2)(N^{2} + 23N + 60)} {(N + 8)^3} \epsilon_{n}^{2}.
\end{eqnarray}
The expression for $\bar{\gamma}_{\phi^{2} (n)}(u_{n \infty})$ in Eq.(88) 
is the same as the one associated to a scalar theory in the massless limit, 
computed at the fixed point using the renormalization group equation. 
Besides, Eq.(89) corresponds to Eq.(95) for this exponent in the 
orthogonal approximation using the massless method of Ref.\cite{generic2}. 

We can compare the results given in Eqs.(86) and (89) of the exponents 
$\eta_{n}$ and $\nu_{n}$ for generic $n$ with the previous massive 
method recently introduced to treat the $n=2$ isotropic case corresponding to 
$m$-axial Lifshitz points in Ref.\cite{CL} employing the orthogonal 
approximation. Two misprints in the 
expressions for the critical exponents 
$\eta_{2} \equiv \eta_{L4}$ and $\nu_{2} \equiv \nu_{L4}$, namely Eqs.(95) 
and (98) in \cite{CL}, respectively, took place in that paper. A wrong factor 
of 2 occurred in Eq.(95) as the coefficient of the $N$ dependent fraction of 
the $O(\epsilon_{L}^{2})$ contribution. In Eq.(98) an incorrect factor of 
$\frac{1}{4}$ also appeared in the $O(\epsilon_{L}^{2})$ contribution therein. 
We emphasize that the correct expressions for those exponents are given by 
Eqs.(86) and (89) for $n=2$ with $\epsilon_{L}$ in \cite{CL} identified 
with $\epsilon_{2}$ herein.
 
\section{Isotropic integrals in the exact calculation}
In the present section we shall discuss and compute the isotropic integrals 
for arbitrary $n$ in order to calculate the critical exponents in Sec.IX. 
This computation in the present massive method is by far much more complicated 
than its counterpart in the massless framework of Ref.\cite{generic2}. Since 
this setting for the case $n=2$ was not explicitly demonstrated in our 
previous work \cite{CL} (the solution to the integrals were only quoted 
in one of the appendixes to that paper), we take this opportunity to fill 
this gap by working out more complicated cases along the same lines of 
reasoning. Here we shall describe in great detail all the technicalities 
involved in the evaluation of these integrals.

As we are going to see, the two- and three-loop contributions for the 
two-point functions represented by the integrals $I_{3}^{'}$ and $I_{5}^{'}$, 
respectively, present no problem in their calculation. In fact, they can be 
solved in terms of the expected poles and regular terms accompanied by 
multiparametric integrals which do not need to be computed explicitly, but 
cancell out in the renormalization algorithm. The issue is the computation of 
one- and two-loop contributions to the (four-point vertex part) coupling 
constant. 

We start by considering the integral $I_{2}$. Since it always appears as a 
subdiagram in higher loop contributions of arbitrary $1PI$ vertex parts, we 
first attempt to compute it at nonvanishing external momenta. In this example 
we shall have a precise idea of the difficulty in this computation. 

At nonvanishing external momenta $K$, $I_{2}$ is given by
\begin{equation}
I_2 =  \int \frac{d^{m_{n}}k}{\bigl((k+K)^{2})^{n} + 1 \bigr)[(k^{2})^{n} + 1]}\;\;\;,
\end{equation}
From elementary complex algebra, we can use the identity 
$(k^{2})^{n} + 1 = (k^{2} - r_{1})(k^{2} - r_{2})...(k^{2} - r_{n})$, where 
$r_{l}$ is the $lth$ complex root of the equation $(k^{2})^{n} + 1 = 0$.  

To see this procedure at work, we apply it to the simplest nontrivial 
case which occurs for $n=2$. In that case, using Feynman 
parameters, the propagator can be written as 
\begin{equation}
\frac{1}{(k^{2}-r_{1})(k^{2}-r_{2})} = \Gamma(2) \int_{0}^{1} 
\frac{dx}{(k^{2} + m_{x}^{2})^{2}},
\end{equation}
where $m_{x}^{2}= (r_{2} - r_{1})x - r_{2}$. In particular, using Feynman 
parameters to fold denominators with different roots amounts to write the 
last expression for $m_{x}^{2}$ in different ways as functions of $r_{1}$ 
and $r_{2}$. Analogous remarks are valid for multiple roots. The final answer 
for the specific integral we are interested in is naturally independent of 
these maneuvers. For the case $n=3$, a similar reasoning leads to the 
following representation for the propagator
\begin{equation}
\frac{1}{(k^{2}-r_{1})(k^{2}-r_{2})(k^{2}-r_{3})} = \Gamma(3) \int_{0}^{1}
\int_{0}^{1} dx_{1} dx_{2} x_{2} 
\frac{1}{(k^{2} + m_{x_{2}}^{2})^{3}},
\end{equation}
with $m_{x_{2}}^{2}= (m_{x_{1}}^{2} + r_{3})x_{2} - r_{3}$ and so on. For 
arbitrary positive integer $n$, we find by the same token the 
following result:
\begin{equation}
\frac{1}{(k^{2}-r_{1})(k^{2}-r_{2})...(k^{2}-r_{n})} = \Gamma(n) 
\int_{0}^{1}...\int_{0}^{1} dx_{1} dx_{2} x_{2}...dx_{n-1} x_{n-1}^{n-2} 
\frac{1}{(k^{2} + m_{x_{n-1}}^{2})^{n}},
\end{equation}
where $m_{x_{n-1}}^{2}= (m_{x_{n-2}}^{2} + r_{n})x_{n-1} - r_{n}$. Now let 
us insert Eq.(93) inside Eq.(90). Consequently $I_{2}$ becomes
\begin{eqnarray}
I_{2}=&& \Gamma(n)^{2} \int_{0}^{1}...\int_{0}^{1} dx_{1} dx_{2} x_{2}...dx_{n-1} x_{n-1}^{n-2}   dy_{1} dy_{2} y_{2}...dy_{n-1} y_{n-1}^{n-2}  \nonumber\\
&& \times\;\; \int d^{m_{n}}k \frac{1}{((k+K)^{2} + m_{x_{n-1}}^{2})^{n} 
(k^{2} + m_{y_{n-1}}^{2})^{n}}.
\end{eqnarray}
We then utilize another Feynman parameter, say $t$, in order to fold the 
two denominators. Integrating over the loop 
momenta \cite{generic2}, we get an angular factor $S_{m_{n}}$ which can be 
discarded/omitted through the redefinition of the coupling constant just as 
before. The reader should be warned that we shall use this fact in all loop 
integrals henceforward. Performing the $\epsilon_{n}$-expansion of the 
prefactors (which are simple Gamma functions), $I_{2}$ turns out to be
\begin{eqnarray}
I_{2}=&& \frac{\Gamma(2n)}{\epsilon_{n}}[1 + \frac{\epsilon_{n}}{2}(\psi(1)- \psi(2n))] \int_{0}^{1}...\int_{0}^{1} dx_{1} dx_{2} x_{2}...dx_{n-1} x_{n-1}^{n-2}   dy_{1} dy_{2} y_{2}...dy_{n-1} y_{n-1}^{n-2}  \nonumber\\
&& \times\; dt[t(1-t)]^{n-1} [t(1-t)K^{2} 
+ (m_{y_{n-1}}^{2} - m_{x_{n-1}}^{2})t + m_{x_{n-1}}^{2}]^{\frac{-\epsilon_{n}}{2}},
\end{eqnarray} 
where $\psi(z)= \frac{d ln \Gamma(z)}{dz}$ is the digamma function. One could 
think that matters get simple to grasp if we set $K=0$ (which is actually the 
symmetry point we shall be concerned in this massive setting), namely 
\begin{eqnarray}
I_{2SP}=&& \frac{\Gamma(2n)}{\epsilon_{n}}[1 + \frac{\epsilon_{n}}{2}(\psi(1)- \psi(2n))] \int_{0}^{1}...\int_{0}^{1} dx_{1} dx_{2} x_{2}...dx_{n-1} x_{n-1}^{n-2}   dy_{1} dy_{2} y_{2}...dy_{n-1} y_{n-1}^{n-2}  \nonumber\\
&& \times\; dt[t(1-t)]^{n-1} [(m_{y_{n-1}}^{2} - m_{x_{n-1}}^{2})t + m_{x_{n-1}}^{2}]^{\frac{-\epsilon_{n}}{2}},
\end{eqnarray} 
but this is not so. The reason for the difficulty even for $K^{2}=0$ is that 
the complex roots of the polynomial in $k$ corresponding to the propagator 
depends explicitly on the value of $n$: we can 
not just go on, within this method, without specifying the value of $n$. This 
is the main obstruction to find the result of the integral in terms of a pole 
in $\epsilon_{n}$ together with a simple regular term for general $n$. 

The simplest way to find the values of the integral for arbitrary $n$ is 
trying to discover a recurrence formula for this integral by analyzing the 
cases with fixed $n$ at the above symmetry point. Two cases are already known. 
The case $n=1$ which corresponds to the standard quadratic $\phi^{4}$ theory, 
whose integral is given by 
$I_{2SP}= \frac{1}{\epsilon_{1}}(1 - \frac{\epsilon_{1}}{2})$. The case $n=2$, 
previously derived in Ref.\cite{CL} is given by 
$I_{2SP}= \frac{1}{\epsilon_{2}}(1 - \frac{\epsilon_{2}}{4})$. It would be 
interesting to compute this integral for higher values of $n$ in order to see 
if the method can be trusted for arbitrary $n$, provided the value of $n$ is 
fixed.   

With this idea in mind, let us compute the integral for $n=3$ in the first 
place. The three complex roots are $r_{1}= -1$, $r_{2}=\frac{1}{2} + i 
\frac{\sqrt{3}}{2} \equiv A$ and $r_{3}=A^{*}$. We find useful to melt 
firstly the contributions of $r_{2},r_{3}$ and folding the resulting 
expression with $r_{1}$ term afterward. Expansion of the digamma 
functions in the prefactors, we find 
\begin{eqnarray}
I_{2SP}=&& \frac{120}{\epsilon_{3}}[1 - \frac{137}{120}\epsilon_{3}] 
\int_{0}^{1}dx \int_{0}^{1}dz\int_{0}^{1}dy y  
\int_{0}^{1} dw w \int_{0}^{1} dt[t(1-t)]^{2}\nonumber\\
&& \times\; [(i \sqrt{3}x - \frac{3}{2} - i \frac{\sqrt{3}}{2})y(1-t) 
+  (i \sqrt{3}z - \frac{3}{2} - i \frac{\sqrt{3}}{2})wt + 1]^{\frac{-\epsilon_{3}}{2}}.
\end{eqnarray} 
When we perform the parametric integrals, we set $\epsilon_{3}=0$ in the 
powers of the resulting complex numbers which remain. The above expression 
is thus equal to 
\begin{equation}
I_{2SP}= \frac{1}{\epsilon_{3}}(1 - \frac{\epsilon_{3}}{6}).
\end{equation}
For good measure, let us analyze the case $n=4$. Since this case is more 
involved and the calculations are rather long (though straightforward), 
we begin with the most basic facts. The complex roots are 
$r_{1}= \frac{1}{2} + i \frac{\sqrt{3}}{2} \equiv B$, $r_{2}= - B^{*}$, 
$r_{3}= -B$ and $r_{4}=B^{*}$. The propagator can be written 
as
\begin{equation}
\frac{1}{(k^{2})^{4} + 1} = \Gamma(4)\int_{0}^{1} dx dy y dz z^{2} 
\frac{1}{(k^{2} + m_{z}^{2})^{4}} ,
\end{equation}
where $m_{z}^{2}= ([(2x-1)B^{*} - B]y + 2B)z - B$. After using the Feynman 
parameters, the integral $I_{2SP}$ then becomes
\begin{equation}
I_{2SP}= \Gamma(4)^{2} \int_{0}^{1} dx dw dy y dr r dz z^{2} ds s^{2} 
\int \frac{d^{m_{4}}k}{(k^{2} + m_{z}^{2})^{4} (k^{2} + m_{s}^{2})^{4}},
\end{equation}
with $m_{s}^{2}= ([(2w-1)B^{*} - B]r + 2B)s - B$. Using another Feynman 
parameter $t$, integrating over the momenta and absorbing the angular factor 
$S_{m_{4}}$ in the redefinition of the coupling constant, we obtain
\begin{eqnarray}
I_{2SP}=&& \frac{\Gamma(8)}{\epsilon_{4}}[1 - \frac{363}{280} \epsilon_{4}] 
\int_{0}^{1}...\int_{0}^{1} dx dw dy y dr r dz z^{2} ds s^{2} dt [t(1-t)]^{3}  \nonumber\\
&& \times\; [(m_{s}^{2}- m_{z}^{2})t + m_{z}^{2}]^{\frac{-\epsilon_{4}}{2}}.
\end{eqnarray}
We are thus left with the task of calculating a large number of elementary 
integrals, which makes the whole process a tedious one. Obviously, the order 
of the integrations chosen was the following: integrate over $x$, $w$, $y$, 
$r$, $z$, $s$ and $t$. In the end of the day, set $\epsilon_{4}=0$ in the 
powers of the remaining complex numbers to get to the result
\begin{equation}
I_{2SP}= \frac{1}{\epsilon_{4}}(1 - \frac{\epsilon_{4}}{8}).
\end{equation} 
From the cases $n=1,2,3,4$, we discover that the method is reliable for 
arbitrary high values of $n$. There is no change of pattern in the 
calculations for different higher values of $n$. Perhaps the only undesirable 
feature is the proliferation of elementary parametric integrals whose number 
increases with increasing $n$. Therefore, for arbitrary $n$ we conclude that
\begin{equation}
I_{2SP}= \frac{1}{\epsilon_{n}}(1 - \frac{\epsilon_{n}}{2n}).
\end{equation}    
We then find out that the orthogonal approximation discussed in Appendix B is 
exact at one-loop level for the massive method.  

Let us turn our attention to the integral $I_{3}$. In a preliminary stage, we 
leave the external momenta arbitrary and in that case it is given by:
\begin{equation}
I_{3} = \int \frac{d^{m_{n}}k_{1} d^{m_{n}}k_{2}}
{\bigl(((k_{1} + k_{2} + K)^{2})^{n} + 1 \bigr)[(k_{1}^{2})^{n}+1][(k_{2}^{2})^{n}+1]}\;\;\;.
\end{equation}
Now, performing the integral over the loop momenta $k_{2}$ by conjugating 
the results expressed in Eqs.(93) and (95), it is easy to 
see that
\begin{eqnarray}
I_{3}&=& \frac{\Gamma(2n) \Gamma(n)}{\epsilon_{n}}[1 
+ \frac{\epsilon_{n}}{2}(\psi(1)- \psi(2n))] \int_{0}^{1}...
\int_{0}^{1} dx_{1} dx_{2} x_{2}...dx_{n-1} x_{n-1}^{n-2} 
dy_{1} dy_{2} y_{2}...dy_{n-1} y_{n-1}^{n-2}  \nonumber\\
&& \times\; dz_{1} dz_{2} z_{2}...dz_{n-1} z_{n-1}^{n-2} 
dt[t(1-t)]^{n -1 -\frac{\epsilon_{n}}{2}} 
\int \frac{d^{m_{n}}k_{1}}
{[k_{1}^{2} + m_{z_{n-1}}^{2}]^{n} [(k_{1}+K)^{2} 
+ m_{t}^{2}]^{\frac{\epsilon_{n}}{2}}},
\end{eqnarray}
where
\begin{equation} 
m_{t}^{2} = \frac{(m_{y_{n-1}}^{2} - m_{x_{n-1}}^{2})t + m_{x_{n-1}}^{2}}
{t(1-t)} \nonumber
\end{equation}
and $m_{z_{n-1}}^{2}= (m_{z_{n-2}}^{2} + r_{n})z_{n-1} - r_{n}$.
We then use another Feynman parameter, say $u$, in order to integrate over 
the loop momenta $k_{1}$ to find 
\begin{eqnarray}
I_{3}&=& \frac{\Gamma(2n) \Gamma(2n - \frac{\epsilon_{n}}{2}) 
\Gamma(-n + \epsilon_{n})}
{2 \Gamma(\frac{\epsilon_{n}}{2}) \epsilon_{n}}[1 
+ \frac{\epsilon_{n}}{2}(\psi(1) - \psi(2n))]
\int_{0}^{1} dx_{1} dx_{2} x_{2}......dx_{n-1} x_{n-1}^{n-2}\nonumber\\ 
&& \times\; dy_{1} dy_{2} y_{2}...dy_{n-1} y_{n-1}^{n-2} 
dz_{1} dz_{2} z_{2}...dz_{n-1} z_{n-1}^{n-2} 
dt[t(1-t)]^{n -1 -\frac{\epsilon_{n}}{2}} \nonumber\\
&& \times\;\; du u^{\frac{\epsilon_{n}}{2} -1}
(1-u)^{n-1}
[u(1-u)K^{2} + (m_{t}^{2} - m_{z_{n-1}}^{2})u 
+ m_{z_{n-1}}^{2}]^{n - \epsilon_{n}} \;\;.
\end{eqnarray}
From this expression we can compute the derivative at zero external momenta, 
i.e., $I_{3}^{'}= \frac{\partial I_{3}}{\partial K^{2n}}|_{K^{2n}=0}$. First, 
we expand the $\Gamma$-functions in the prefactor using the 
appropriate identities for these functions. Second, after taking the 
derivative and setting the external momenta to zero, the first term in the 
last bracket contributes to the integral over $u$, whereas the powers of the 
remaining parameters are expanded in $\epsilon_{n}$ through the elementary 
identity $a^{-\epsilon_{n}}(u,t,z_{n-1})= 1 - \epsilon_{n} \;ln\;a(u,t,z_{n-1})$. 
Thirdly, the identity \cite{GR}
\begin{equation}
\int_{0}^{1} dx x^{\mu -1}(1-x^{r})^{\nu - 1} lnx = \frac{1}{r^{2}} 
B(\frac{\mu}{r},\nu)[\psi(\frac{\mu}{r}) - \psi(\frac{\mu}{r} + \nu),
\end{equation}
where B(x,y) is the Euler beta function, will be useful to our purposes. It 
allows to separate the pole and a regular term by solving all the parametric 
integrals together with a regular term which depends on only one 
multiparametric integral. Collecting together this set of steps, 
$I_{3}^{'}$ turns out to be
\begin{eqnarray}
I_{3}^{'}= &&\frac{(-1)^{n} \Gamma(2n)^{2}}
{4 \Gamma(n+1) \Gamma(3n) \epsilon_{n}} [1 + \epsilon_{n}(\frac{-3}{4} 
- \sum_{p=3}^{3n-1} \frac{1}{2p} + \sum_{p=1}^{n-1} \frac{1}{2p})]\nonumber\\
&& + (-1)^{n+1} \frac{\Gamma(2n)^{2}}{4 \Gamma(n+1)} H',
\end{eqnarray}
where the remaining multiparametric integral $H'$ is given by
\begin{eqnarray}
H'&=& \int_{0}^{1} dx_{1} dx_{2} x_{2}......dx_{n-1} x_{n-1}^{n-2} 
dy_{1} dy_{2} y_{2}...dy_{n-1} y_{n-1}^{n-2} 
dz_{1} dz_{2} z_{2}...dz_{n-1} z_{n-1}^{n-2} dt[t(1-t)]^{n-1}\nonumber\\ 
&&\times\; du u^{n-1}(1-u)^{2n-1} 
ln\bigr(\bigl[\frac{((m_{y_{n-1}}^{2} - m_{x_{n-1}}^{2})t + m_{x_{n-1}}^{2})}
{t(1-t)} - m_{z_{n-1}}^{2} \bigr] u + m_{z_{n-1}}^{2} \bigr) \;\;.
\end{eqnarray} 
The above integral is the generalization of the integral $H$ Eq.(C5) from 
Ref.\cite{CL} appearing in the exact computation of the analogous two-point 
diagram for $m$-axial Lifshitz points.

After this discussion, it is straightforward to perform the three-loop 
integral contributing to the two-point function, namely
\begin{equation}
I_{5} = \int \frac{d^{m_{n}}k_{1} d^{m_{n}}k_{2}d^{m_{n}}k_{3}}
{\bigl(((k_{1} + k_{2} + K)^{2})^{n} + 1 \bigr) \bigl(((k_{1} + k_{3} + K)^{2})^{n} + 1 \bigr)((k_{1}^{2})^{n} + 1)  ((k_{2}^{2})^{n}+ 1) ((k_{3}^{2})^{n} + 1) }\;\;\;.
\end{equation}
Note that the internal subdiagram of the four-point function appears 
quadratically in the above integral. Using Eq.(93) in conjunction with 
Eq.(95), we get to the following intermediate result
\begin{eqnarray}
I_{5}&=& \frac{\Gamma(2n) \Gamma(n)}{\epsilon_{n}^{2}}[1 
+ \epsilon_{n} (\psi(1)- \psi(2n))] \int_{0}^{1}...
\int_{0}^{1} dx_{1} dx_{2} x_{2}...dx_{n-1} x_{n-1}^{n-2} 
dy_{1} dy_{2} y_{2}...dy_{n-1} y_{n-1}^{n-2}  \nonumber\\
&& \times\; dz_{1} dz_{2} z_{2}...dz_{n-1} z_{n-1}^{n-2} 
dt[t(1-t)]^{n -1- \epsilon_{n}} 
\int \frac{d^{m_{n}}k_{1}}
{[k_{1}^{2} + m_{z_{n-1}}^{2}]^{n} [(k_{1}+K)^{2} 
+ m_{t}^{2}]^{\epsilon_{n}}}.
\end{eqnarray}
Proceeding as before, integrate over the remaining loop momenta utilizing 
another Feynman parameter. This leads to
\begin{eqnarray}
I_{5}&=& \frac{\Gamma(2n) \Gamma(2n - \frac{\epsilon_{n}}{2}) 
\Gamma(-n + \frac{3 \epsilon_{n}}{2})}
{2 \Gamma(\epsilon_{n}) \epsilon_{n}^{2}}[1 
+ \epsilon_{n}(\psi(1) - \psi(2n))]
\int_{0}^{1} dx_{1} dx_{2} x_{2}......dx_{n-1} x_{n-1}^{n-2}\nonumber\\ 
&& \times\; dy_{1} dy_{2} y_{2}...dy_{n-1} y_{n-1}^{n-2} 
dz_{1} dz_{2} z_{2}...dz_{n-1} z_{n-1}^{n-2} 
dt[t(1-t)]^{n -1 -\epsilon_{n}} \nonumber\\
&& \times\;\; du u^{\epsilon_{n} -1}
(1-u)^{n-1}
[u(1-u)K^{2} + (m_{t}^{2} - m_{z_{n-1}}^{2})u 
+ m_{z_{n-1}}^{2}]^{n - \frac{3 \epsilon_{n}}{2}} \;\;.
\end{eqnarray}
We then compute 
$I_{5}^{'}= \frac{\partial I_{5}}{\partial K^{2n}}|_{K^{2n}=0}$. Taking the 
derivative, performing the expansion of the $\Gamma$-functions, using 
$u^{\epsilon_{n}}= 1 + \epsilon_{n} lnu$ along with a similar expansion 
for the brackets (naturally taken at $K^{2n}=0$) and employing the identity 
(107), the final result for the integral is
\begin{eqnarray}
I_{5}^{'}= &&\frac{(-1)^{n} \Gamma(2n)^{2}}
{3 \Gamma(n+1) \Gamma(3n) \epsilon_{n}^{2}} [1 + \epsilon_{n}(-1 
- \sum_{p=3}^{3n-1} \frac{1}{p} + \sum_{p=1}^{n-1} \frac{1}{2p} 
+ \sum_{p=1}^{2n-1} \frac{1}{2p})]\nonumber\\
&& + (-1)^{n+1} \frac{\Gamma(2n)^{2}}{3 \Gamma(n+1)\epsilon_{n}} H',
\end{eqnarray}
where $H'$ is given by eq.(109).

Finally, let us compute $I_{4}$ at zero external momenta, namely
\begin{eqnarray}
I_{4}\;\; =&& \int \frac{d^{m_{n}}k_{1}d^{m_{n}}k_{2}}
{[(k_{1}^{2})^{n}+1]^{2} 
\left((k_{2}^{2})^{n} + 1 \right) \bigl(((k_{1} + k_{2})^{2})^{n} + 1 \bigl)}\;\;.
\end{eqnarray}
The same kind of limitation taking place for $I_{2}$ occurs for $I_{4}$: the 
lack of a solution for general $n$ due to the fact that the roots of the 
polynomial in the momenta in our new representation of the propagator depends 
explicitly on $n$. To see this, integrate first over the loop momenta $k_{2}$ 
just as explained above, that is, use Feynman parameters to rewrite $I_{4}$ 
as 
\begin{eqnarray}
I_{4}&=& \frac{\Gamma(2n) \Gamma(n)}{\epsilon_{n}}[1 
+ \frac{\epsilon_{n}}{2}(\psi(1)- \psi(2n))] \int_{0}^{1}...
\int_{0}^{1} dx_{1} dx_{2} x_{2}...dx_{n-1} x_{n-1}^{n-2} 
dy_{1} dy_{2} y_{2}...dy_{n-1} y_{n-1}^{n-2}  \nonumber\\
&& \times\; dt[t(1-t)]^{n -1 -\frac{\epsilon_{n}}{2}} 
\int \frac{d^{m_{n}}k_{1}}
{[(k_{1}^{2})^{n} + 1]^{2} [(k_{1})^{2} 
+ m_{t}^{2}]^{\frac{\epsilon_{n}}{2}}}\;\;.
\end{eqnarray}
If we use new sets of Feynman parameters $z_{i}$ to work out the first 
propagator, this operation produces the expression
\begin{eqnarray}
I_{4}&=& \frac{\Gamma(2n) \Gamma(n)}{\epsilon_{n}}[1 
+ \frac{\epsilon_{n}}{2}(\psi(1)- \psi(2n))] \int_{0}^{1}...
\int_{0}^{1} dx_{1} dx_{2} x_{2}...dx_{n-1} x_{n-1}^{n-2} 
dy_{1} dy_{2} y_{2}...dy_{n-1} y_{n-1}^{n-2}  \nonumber\\
&& \times\; dt[t(1-t)]^{n -1 -\frac{\epsilon_{n}}{2}}
dz_{1}z_{1}(1-z_{1}) dz_{2} z_{2}^{3}(1-z_{2})...dz_{n-1} z_{n-1}^{2n-3}
(1-z_{n-1})\nonumber\\
&&\int \frac{d^{m_{n}}k_{1}}
{[k_{1}^{2} + m_{z_{n-1}}^{2}]^{2n} [(k_{1})^{2} 
+ m_{t}^{2}]^{\frac{\epsilon_{n}}{2}}}\;\;.
\end{eqnarray}
Performing another integral over the remaining loop momenta the integral 
can be entirely represented by parametric integrals, that is
\begin{eqnarray}
I_{4}&=& \frac{\Gamma(2n) \Gamma(2n - \frac{\epsilon_{n}}{2}) 
\Gamma(\epsilon_{n})}
{2 \Gamma(\frac{\epsilon_{n}}{2}) \epsilon_{n}}[1 
+ \frac{\epsilon_{n}}{2}(\psi(1) - \psi(2n))]
\int_{0}^{1} dx_{1} dx_{2} x_{2}......dx_{n-1} x_{n-1}^{n-2}\nonumber\\ 
&& \times\; dy_{1} dy_{2} y_{2}...dy_{n-1} y_{n-1}^{n-2} 
dz_{1}z_{1}(1-z_{1}) dz_{2} z_{2}^{3}(1-z_{2})...dz_{n-1} z_{n-1}^{2n-3}
(1-z_{n-1})\nonumber\\
&& \times\;\; dt[t(1-t)]^{n -1 -\frac{\epsilon_{n}}{2}} 
du u^{\frac{\epsilon_{n}}{2} -1}
(1-u)^{2n-1}
[(m_{t}^{2} - m_{z_{n-1}}^{2})u 
+ m_{z_{n-1}}^{2}]^{n - \epsilon_{n}} \;\;.
\end{eqnarray}
A standard procedure is to compute the integrand at $u=0$: by summing and 
subtracting the integrand at this value, the difference between the integral 
at $u \neq 0$ and at $u=0$ is higher order in $\epsilon_{n}$ and can be 
neglected \cite{new22,PS}. We emphasize the factor $m_{z_{n-1}}^{2}$ depends upon 
the variables $z_{i}$. Therefore, we can integrate over the other parameters 
and performing the $\epsilon_{n}$-expansion of the $\Gamma$-functions, we obtain the following expression:
\begin{eqnarray}
I_{4}&=& \frac{\Gamma(2n)}
{2 \epsilon_{n}^{2}}[1 
+ \epsilon_{n}(\frac{3}{2} \psi(1) - \frac{1}{2}\psi(2n) - \psi(n))]\nonumber\\ && \times\; \int_{0}^{1} dz_{1}z_{1}(1-z_{1}) dz_{2} z_{2}^{3}(1-z_{2})...
dz_{n-1} z_{n-1}^{2n-3}(1-z_{n-1})(m_{z_{n-1}}^{2})^{- \epsilon_{n}} \;\;.
\end{eqnarray} 
In order to proceed from this point, we must specify again the values of $n$ 
in order to determine the roots explicitly in order to solve the remaining 
parametric integrals. We already know this integral for $n=1,2$, namely
\begin{equation}
I_{4}= \frac{1}{2 \epsilon_{1}^{2}}\bigr(1-\frac{\epsilon_{1}}{2}\bigr),
\end{equation}  
\begin{equation}
I_{4}= \frac{1}{2 \epsilon_{2}^{2}}\bigr(1-\frac{7 \epsilon_{2}}{12}\bigr).
\end{equation} 
The cases $n=3,4$ can be worked out just as we did in the one-loop case. With 
the resources furnished so far, it is not difficult to obtain the following 
results 
\begin{equation}
I_{4}= \frac{1}{2 \epsilon_{3}^{2}}\bigr(1-\frac{83\epsilon_{3}}{120}\bigr),
\end{equation}
\begin{equation}
I_{4}= \frac{1}{2 \epsilon_{4}^{2}}\bigr(1-\frac{661 \epsilon_{4}}{840}\bigr).
\end{equation}
The value of $I_{4}$ for general $n$ which correctly reduces to the above 
particular cases is then given by
\begin{equation}
I_{4}= \frac{1}{2 \epsilon_{n}^{2}}
\bigr(1 + \epsilon_{n}(D(n)-\frac{1}{n})\bigr).
\end{equation}
where $D(n)= \frac{1}{2} \psi(1) - \psi(n) + \frac{1}{2} \psi(2n)$.

This completes our task of calculating explicitly the exact isotropic 
integrals in the massive case \cite{PhD} necessary to compute the critical 
exponents. This aim shall be tackled in the next section.

\section{Isotropic exponents in the exact computation}
Now we apply the normalization conditions of the massive theory in pretty 
much the same way as worked out in the anisotropic and isotropic cases. In 
other words, the algorithm to determine the critical exponents is the same, 
but now we have to replace the values of the integrals calculated in last 
section.  

The coefficients of the several renormalization functions can be easily 
found. Indeed, using the Wilson functions appropriate to the isotropic 
cases the cofficients have the following expressions:
\begin{subequations}
\begin{eqnarray}
&& a_{1 n} = \frac{N+8}{6 \epsilon_{n}}[1 - \frac{1}{2n} \epsilon_{n}] ,\\
&& a_{2 n} = (\frac{N+8}{6 \epsilon_{n}})^{2}
- [\frac{N^{2} + 26N + 108}{36n \epsilon_{n}}] 
+ \frac{5N+22}{18\epsilon_{n}}\bigl(\frac{1}{n} - D(n)\bigr) \nonumber\\
&& - (-1)^{n} \frac{\Gamma(2n)^{2} (N+2)}{36 \Gamma(n+1) 
\Gamma(3n) \epsilon_{n}},\\
&& b_{2 n} = (-1)^{n} \frac{\Gamma(2n)^{2}(N+2)}{72 \Gamma(n+1) \Gamma(3n) 
\epsilon_{n}}[1 + \epsilon_{n}(- \frac{3}{4} + \sum_{p=1}^{n-1} \frac{1}{2p} 
- \sum_{p=3}^{3n-1} \frac{1}{2p} - \Gamma(3n) H')] , \\
&& b_{3 n} = (-1)^{n}\frac{\Gamma(2n)^{2}(N+2)(N+8)}
{648 \Gamma(3n)\epsilon_{n}^{2}}\bigl(1 + \epsilon_{n}(-\frac{1}{4} 
+ \sum_{p=1}^{n-1} \frac{1}{2p} - \sum_{p=1}^{2n-1} \frac{1}{p}\nonumber\\
&& + \sum_{p=3}^{3n-1} \frac{1}{2p} -\frac{3}{2n})\bigr), \\
&& c_{1 n} = \frac{(N+2)}{6 \epsilon_{n}}[1 - \frac{1}{2n} \epsilon_{n}], \\
&& c_{2 n} = \frac{(N+2)}{6 \epsilon_{n}^{2}}\bigl[\frac{(N+5)}{6}
\bigl(1 - \epsilon_{n} \frac{1}{n}\bigr) -\frac{1}{2} D(n) \epsilon_{n}\bigr].
\end{eqnarray}
\end{subequations}   
In practice these results determine the several renormalization functions. They are given by:
\begin{subequations}
\begin{eqnarray}
&&\beta_{n}(u_{n})= -u_{n}\bigl[ \epsilon_{n} - \frac{N+8}{6}
\bigl(1 - \frac{\epsilon_{n}}{2n}\bigr) u_{n} + \bigl(\frac{5N+22}{9}D(n) 
+ (-1)^{n} \frac{\Gamma(2n)^{2} (N+2)}{18 \Gamma(n+1) \Gamma(3n)} \bigr)
\nonumber \\ 
&& \;\;\; \times \;\;\; u_{n}^{2} \bigr], \\
&&\gamma_{\phi(n)}(u(n)) = (-1)^{n+1} \frac{\Gamma(2n)^{2}(N+2)}
{36 \Gamma(n+1) \Gamma(3n) \epsilon_{n}} un^{2} 
\bigl[1+\epsilon_{n}(- \frac{3}{4} + \sum_{p=1}^{n-1} \frac{1}{2p} 
- \sum_{p=3}^{3n-1} \frac{1}{2p} - \Gamma(3n) H')\nonumber\\ 
&& \; + \frac{N+8}{6}(\frac{1}{2} - \sum_{p=2}^{2n-1} \frac{1}{2p} 
+ \sum_{p=3}^{3n-1} \frac{1}{2p} - \frac{1}{n} + \Gamma(3n) H') u_{n}\bigr],\\
&& \bar{\gamma}_{\phi^{2} (n)}(u_{n})= \frac{N+2}{6} u_{n} \bigl[1 
- \frac{\epsilon_{n}}{n} - D(n) u_{n} \bigr].
\end{eqnarray}
\end{subequations}
The fixed points $u_{n \infty}$ are obtained from the eigenvalue conditions 
$\beta_{n}(u_{n \infty}) = 0$, which yield
\begin{equation}
u_{n \infty}= \frac{6}{N+8}\epsilon_{n}\bigl[1 
+ \epsilon_{n} \bigl(\bigl[(20N+88)D(n) + (-1)^{n} 
\frac{\Gamma(2n)^{2} (2N+4)}{\Gamma(n+1) \Gamma(3n)} \bigr]\frac{1}{(N+8)^{2}} 
+ \frac{1}{2n} \bigr) \bigr].
\end{equation}
It should be pointed out that neither the coefficients/renormalization 
functions nor the fixed points are equal to those obtained in Sec.VII of 
Ref.\cite{generic2} for the massless theory: the normalization conditions are 
different and the difference is due to the nonuniversal feature of these 
functions away from the fixed points. Incidentally, we make the observation 
that even the fixed points are not universal but actually vary with the 
renormalization scheme employed.  

Replacing the fixed points back into $\gamma_{\phi(n)}(u_{n})$, the result is 
just the anomalous dimension $\eta_{n}= \gamma_{\phi(n)}(u_{n \infty})$, 
which is given by
\begin{eqnarray}
\eta_{n} = (-1)^{n+1} 
\frac{(N+2)\Gamma(2n)^{2}}{(N+8)^{2} \Gamma(n+1) \Gamma(3n)} \epsilon_{n}^{2} 
+(-1)^{n+1}\frac{(N+2)\Gamma(2n)^{2} F(N,n)}{(N+8)^{2} \Gamma(n+1) \Gamma(3n)} 
\epsilon_{n}^{3}\;\;,
\end{eqnarray}
where 
\begin{eqnarray}
&&F(N,n)= \bigl[ \bigl((-1)^{n} \frac{\Gamma(2n)^{2} (4N+8)}{\Gamma(n+1) 
\Gamma(3n)} + (40N+176)D(n) \bigr) \frac{1}{(N+8)^{2}}\nonumber\\
&&- \sum_{p=1}^{2n-1} \frac{1}{p} + \frac{1}{2} \sum_{p=1}^{n-1} 
\frac{1}{p} + \frac{1}{2} \sum_{p=1}^{3n-1} \frac{1}{p}\bigr].
\end{eqnarray}
Comparing this expression with Eq.(111) from Ref.\cite{generic2}, we 
detected a misprint therein: an extra factor of $\frac{-3}{4}$ appearing 
there should be disconsidered. The correct expression  for $F(N,n)$ is given 
by the last equation shown above. 

Note that the integral $H'$ cancelled out in the computation of $\eta_{n}$. 
Finally let us compute the exponent $\nu_{n}$. First, we write 
$\bar{\gamma}_{\phi^{2} (n)}(u_{n})$ at the fixed point, and then use the 
scaling relation among  $\nu_{n}, \bar{\gamma}_{\phi^{2} (n)}(u_{n \infty})$ 
and $\eta_{n}$ to get the following result:
\begin{eqnarray}
&& \nu_{n} = \frac{1}{2n} + \frac{(N+2)}{4n^{2}(N+8)}\epsilon_{n} 
+ \frac{(N+2)}{4n^{2}(N+8)^{3}} \epsilon_{n}^{2}
[(-1)^{n} (N-4) \frac{\Gamma(2n)^{2}}{\Gamma(n+1) \Gamma(3n)} \nonumber\\
&& + \frac{(N+2)(N+8)}{2n}  + (14N+40)D(n)].
\end{eqnarray}
These universal quantities are the same as those obtained in 
Ref.\cite{generic2} in the zero mass approach. The other critical 
exponents can be obtained by scaling relations from the two exponents 
above and are also the same as those calculated from the zero mass setting. 
A rather interesting point is the following: while the exponents in the zero 
mass limit are obtained analyzing the behavior of the vertex parts in the 
infrared, the exponents here are obtained in the ultraviolet regime since the 
fixed points $u_{n \infty}$ are ultraviolet fixed points. We thus have 
complete equivalence of the zero mass theory renormalized at nonzero external 
momenta with the massive theory renormalized at vanishing external momenta. 
Universality is obeyed as expected. 

\section{Final Comments}
\par The introduction of a massive method to calculating critical exponents 
of generic competing systems is another step forward to a better 
comprehension of Lifshitz criticalities for several reasons. First, the 
massive method along with the Callan-Symanzik-Lifshitz equations is 
appropriate to proving the multiplicative renormalizability to all orders 
in perturbation theory as demonstrated herein. This proof for generic 
competing systems is a nontrivial generalization of that already performed 
for $m$-axial Lifshitz points \cite{CL}. Besides, the massive framework with 
its complete equivalence to the zero mass treatment reflected in identical 
critical exponents with those in Refs.\cite{generic2} whether we use 
approximations for the appropriate cases or approach the calculations exactly 
is another exact manifestation of universality. Let us discuss some properties 
of anisotropic and isotropic cases separately and how they can show up in 
other field-theoretic contexts.

\par It is worthy to emphasize some especial features that occur in both 
massless and massive renormalized perturbation theories constructed to 
describe the 
generic higher character Lifshitz universality classes. As we have discussed, 
the computations in the massive theory is much more elaborate than in the 
massless method. Even though the Wilson functions determining the renormalized 
theory are explicitly different in both cases, they converge to the same value 
at the respective fixed points, which make them responsible for the same 
values of the critical exponents (and all universal quantities) in either 
scheme. This is not the only difference: while the massless $1PI$ vertex parts 
are scale invariant at the infrared attractive fixed point $u_{n}^{*}$, their 
counterparts in the massive theory are scale invariant precisely at the 
ultraviolet nonattractive fixed point $u_{n \infty}$. The same remark was 
already pointed out in the $m$-axial Lifshitz universality \cite{CL} 
following closely the behavior of the pure $\phi^{4}$ noncompeting case 
argument of BLZ \cite{new25}. The several independent 
mass scales implementing independent renormalization group transformations in 
each type of competing subspace correspond to the several correlation lengths 
appearing in each modulated ordered phase in the CECI model.       
  
\par We can consider direct applications of the present formalism in the first 
place. We can use the method to compute critical amplitude ratios of various 
thermodynamic potentials (as well as other universal observables constructed 
from the renormalized $1PI$ vertex functions) above an below the Lifshitz 
temperature. Some uniaxial $m$-fold Lifshitz critical amplitudes have already 
been calculated at one-loop level \cite{Leite,L1}. Even in this simpler case, 
not all amplitudes were computed and much more can be done. The motivation is 
to compare the results with new materials which might present Lifshitz points 
of generic higher character involving several independent length scales. 
Indeed, the comparison of the specific heat amplitude ratio of Ref.\cite{L1} 
with the one obtained experimentally for the material $MnP$ \cite{BBO} yielded 
a remarkable agreement between the two results. Our hope is that the present 
method can help to unveil all the critical amplitude ratios, therefore 
generalizing the analysis for the $m$-axial Lifshitz with just two independent 
length scales. Moreover, we hope that numerical calculations such as those 
studied in high temperature series for the uniaxial Lifshitz case at fixed 
values of the space dimension and number of components of the order 
parameter \cite{PB} shall be put forth in order to improve our 
understanding of the universality classes of the most general competing 
systems.

\par Beyond the problem of critical phenomena occurring in generic competing 
critical systems, the mathematical apparatus developed in the present 
paper might be useful to attack the perturbative aspects of many quantum 
field theories with higher derivatives that have been studied recently. It has 
been pointed out that field theories with higher space derivatives and some 
sort of mechanism like the Lifshitz condition which suppresses the second 
space derivatives eliminate the lack of unitarity and produces a well behaved 
theory without ghosts \cite{Anselmi}. It shares a similar renormalization 
group treatment: the time scale behaves in the same way as the noncompeting 
direction since it is quadratic in the derivatives, whereas the space 
directions have different scaling dimensions analogous to the Lifshitz 
competing directions.  These Lorentz violating models have also 
been constructed in the cases of gauge theories \cite{Anselmi2} with 
implications in the standard model and the analysis of neutrino masses 
\cite{Anselmi3}. 

\par Although not studied yet within our present investigation, the issue of 
quantum critical behaviors of Lifshitz points might also be worthwhile.
An instigating aspect which appeared recently in the literature is the 
application of these ideas in examinating the quantum criticality of 
membranes, which is analogous to the anisotropic $m$-axial $n=2$ case with 
two length scales \cite{Horava1}. This description corresponds to a new class 
of gravity models. It has been further speculated that the case $n=3$ of third 
character anisotropic Lifshitz points can be studied in this framework and 
explains the basic features of quantum gravity at these Lifshitz points 
\cite{Horava2}. This study adds more ingredients to understand the infrared 
modifications of gravity described by the ghost condensate previously 
introduced in Ref.\cite{AH}, which has a more transparent analogy with the 
Lifshitz CECI model: gravity could have attractive character at ``short'' 
distances, but might develop repulsive properties in the long distance 
limit. 

\par Furthermore, the conjectures put forward in \cite{CL} can be further 
extended to the most general cases of arbitrary independent mass scales: 
each scale is characterized by an independent Compton wavelength. This is 
kind of bizarre, since now more than one mass scale could characterize the 
corresponding ``elementary particle''. This anisotropy could reflect the way 
it is distributed over space(time). In other words, if space(time) would be 
anisotropic (as suggested by these many independent mass scales) this feature 
could leave a mark on particles propagating on it. The connection with the 
model discussed here can be easily viewed in considering a scalar quantum 
field with a Lifshitz-like condition which only keeps higher space 
derivatives and maintains solely second order time derivatives. A Wick 
rotation 
brings this theory to the statistical mechanics form of the Lagrangian 
density Eq.(2). In this ``generalized Lifshitz space'', the Laplacian operator 
is originally defined in $d$ dimensions, but due to distinct types of 
competition axes (caused by forces with alternate signals) as realized 
in the CECI model, only $(d- \sum_{n=2}^{L} m_{n})$ of its components (e.g., 
the time components) remain. The residual information that the field lived in 
$d$-dimensions in the first place before dynamical effects break these 
pattern is encoded in the higher derivative terms in the Lagrangian density. 

\par Minkowski space corresponds to the limit of zero gravity, where quantum 
fields propagate in. Competition between repulsive and attractive components 
of gravity in the large distance limit could then produce a tiny effect, which 
could result in this Lifshitz space. When Wick rotated back, this would 
produce precisely another type of flat space limit where gravity is small 
but has an observable effect on it: Lorentz invariance is broken in the 
bare propagator of the quantum fields, since now the timelike components of 
the momentum is quadratic as before, but the spacelike components have higher 
powers. Choosing independent metrics in each 
competing subspace in the Lagrangian (2), the higher order term corresponding 
to the $m_{n}$ dimensional competing subspace could be written as 
$g_{rs}(\partial_{r})^{n}(\partial_{s})^{n}$, with $r,s=1,...,n$ for 
$g_{rs}=\delta_{rs}$. With this choice, the introduction of masses in 
dimensional reduction using the Siegel method would follow simply as discussed 
for each uniaxial subspace $m_{n}=1$ without any further complication. We 
believe that the perturbative analysis of the examples above mentioned 
incorporating these new ideas might be worthwhile 
after the study described in the present work.  

\par Most of the remarks done in the anisotropic cases can be extended to 
isotropic points. From a realistic point of view, however, it is much more 
difficult to visualize examples in quantum field theories: as pointed out 
before, more than two time derivatives in the Lagrangian density leads to 
trouble with unitarity. We can make remarks concerning the method itself in 
comparison with the zero mass case previously discussed in Ref.\cite{generic2}.

\par The massive integrals computed using the orthogonal approximation are 
very simple and can be computed explicitly for generic number of neighbors 
coupled via competing interactions $n$. On the other hand, the exact massive 
integrals are rather involved, even at the one-loop order, within the proposal 
presented. Just as happens in the massless case, the one-loop massive diagram 
explicit computation with general $n$ is identical to the result using the 
orthogonal approximation. The latter is exact at one loop. Deviations start 
at two-loop order and beyond. What is really remarkable using either the 
massless or the massive approach is that usual systems without competition 
are particular cases of generic isotropic criticalities with $n=1$, i.e., 
they are Lifshitz behaviors of first character. 

\par In summary, we evaluated the critical indices of generic higher character 
Lifshitz points using massive fields along with normalization conditions 
defining the renormalized theories at zero external momenta with many 
independent masses responsible for independent renormalization group 
transformations in each competing subspace. The results turn out to be 
identical to those obtained previously using a massless framework. Thus, 
universality is corroborated once again.

\section{Acknowledgments}
We would like to thank partial financial support by CNPq from Brazil.

\appendix 
\section{Feynman graphs for anisotropic behaviors}

The required diagrams corresponding to one-, two- and three-loop 
integrals in momenta space shall be computed using dimensional regularization 
and Feynman parameters as our main tools in their solution. Solely the 
one-loop integral associated to the four-point $1PI$ vertex part can be solved 
exactly for anisotropic behaviors. Multiloop graphs can be evaluated through 
the generalized orthogonal approximation. The main step is actually quite 
simple: {\it the loop momenta characterizing a certain competition subspace 
in a given bubble (subdiagram) do not mix to all loop momenta not
belonging to that bubble}. In other words, the loop 
momenta in a given subdiagram is orthogonal to all loop/external momenta 
appearing in other subdiagrams. The normalization conditions defined in the 
text indicates that the integrals to be worked out should be determined at 
zero external momenta. The minimal set of Feynman integrals to be solved are 
presented with increasing order in the number of propagators, namely 
the one-loop integral contributing to the four-point function
\begin{equation}
I_2 =  \int \frac{d^{d-\sum_{n=2}^{L} m_{n}}q \Pi_{n=2}^{L} d^{m_{n}}k_{(n)}}
{[\sum_{n=2}^{L}(k_{(n)}^{2})^{n} +
(q)^{2} + 1]^{2}}\;\;\;,
\end{equation}
the two-loop contribution to the two-point function 
$I_{3}'=\frac{\partial I_{3}(P,K_{(n)})}{\partial P^{2}}|_{P=K_{(n)}=0}\\
(= 
\frac{\partial I_{3}(P,K_{(n)})}{\partial K_{(n)}^{2n}}|_{P=K_{(n)}=0})$, 
where $I_{3}(P,K_{(n)})$ is the integral
\begin{eqnarray}
I_{3} =&& \int \frac{d^{d-\sum_{n=2}^{L} m_{n}}{q_{1}}d^{d-\sum_{n=2}^{L} m_{n}}q_{2} \Pi_{n=2}^{L} d^{m_{n}}k_{1 (n)}\Pi_{n=2}^{L} d^{m_{n}}k_{2 (n)}}
{\left( q_{1}^{2} + \sum_{n=2}^{L}(k_{1 (n)}^{2})^{n}+1 \right)
\left( q_{2}^{2} +  \sum_{n=2}^{L}(k_{2 (n)}^{2})^{n}+1 \right)}\nonumber\\
&& \qquad\qquad \times \frac{1}{[(q_{1} + q_{2} + P)^{2} + \sum_{n=2}^{L}((k_{1 (n)} + k_{2 (n)} + K_{(n)})^{2})^{n} +1]}\;\;,
\end{eqnarray}
the two-loop contribution to the coupling constant
\begin{eqnarray}
I_{4} =&& \int \frac{d^{d-\sum_{n=2}^{L} m_{n}}{q_{1}}d^{d-\sum_{n=2}^{L} m_{n}}q_{2}\Pi_{n=2}^{L} d^{m_{n}}k_{1 (n)}\Pi_{n=2}^{L} d^{m_{n}}k_{2 (n)}}
{\left ( q_{1}^{2} + \sum_{n=2}^{L}(k_{1 (n)}^{2})^{n} + 1 \right)^{2}}\nonumber\\
&&\qquad \times \frac{1}
{\left( q_{2}^{2} +  \sum_{n=2}^{L}(k_{2 (n)}^{2})^{n}+ 1 \right)
[(q_{1} + q_{2})^{2} + \sum_{n=2}^{L}\bigl((k_{1 (n)} 
+ k_{2 (n)})^{2}\bigl)^{n} +1 ]}\;\;,
\end{eqnarray}
and finally the three-loop integral 
$ I_{5}'= \frac{\partial I_{3}(P,K_{(n)})}{\partial P^{2}}|_{P=K_{(n)}=0}(= 
\frac{\partial I_{3}(P,K_{(n)})}{\partial K_{(n)}^{2n}}|_{P=K_{(n)}=0})$ 
with $I_{5}(P,K_{(n)})$ representing the graph
\begin{eqnarray}
I_{5} &=&
\int \frac{d^{d-\sum_{n=2}^{L} m_{n}}q_{1} d^{d-\sum_{n=2}^{L} m_{n}}q_{2} d^{d-\sum_{n=2}^{L} m_{n}}q_{3} \Pi_{n=2}^{L} d^{m_{n}}k_{1 (n)}}
{\left( q_{1}^{2} + \sum_{n=2}^{L}(k_{1 (n)}^{2})^{n} + 1 \right)
\left( q_{2}^{2} + \sum_{n=2}^{L}(k_{2 (n)}^{2})^{n} + 1 \right)
\left( q_{3}^{2} + \sum_{n=2}^{L}(k_{3 (n)}^{2})^{n}+ 1 \right)}\nonumber\\
&& \qquad\qquad\qquad \times \frac{\Pi_{n=2}^{L} d^{m_{n}}k_{2 (n)} \Pi_{n=2}^{L} d^{m_{n}}k_{3 (n)}}{[ (q_{1} + q_{2} - P)^{2} + \sum_{n=2}^{L} \bigl((k_{1(n)} + k_{2(n)} - K_{(n)})^{2}\bigr)^{n}+ 1 ]}\nonumber\\
&& \qquad\qquad\qquad \times \frac{1}{[(q_{1} + q_{3} - P)^{2} + \sum_{n=2}^{L}\bigl((k_{1(n)} + k_{3(n)} -
  K_{(n)})^{2}\bigr)^{n}+ 1 ]}.
\end{eqnarray}
We stress that in the expressions for $I_{3}$ and $I_{5}$, $P$ is the external 
momenta perpendicular to the several types of competing axes 
(,i.e., it belongs to the $m_{1}$ subspace) whereas 
$K_{(n)}$ $(n=2,...,L)$ is the external momenta representing the $nth$ 
$m_{n}$-dimensional competing subspace. They are only needed to compute the 
derivative of these integrals, but shall be set to zero after that operation. 

One should keep in mind that whenever a loop integral is performed, the 
geometric angular factor $[S_{(d-\sum_{n=2}^{L}m_{n})} 
\Gamma(2 - \sum_{n=2}^{L}\frac{m_{n}}{2n})(\Pi_{n=2}^{L} 
\frac{S_{m_{n}} \Gamma(\frac{m_{n}}{2n})}{2n})]$ appears but can be omitted 
in the final result of the integrals by a redefinition of the coupling 
constant. Performing this redefinition and applying the mathematical tools 
explained above, we learn that those 
divergent integrals have the following representation in terms of dimensional 
poles corresponding to their $\epsilon_{L}$-expansion values: 
\begin{equation}
I_{2} = \frac{1}{\epsilon_{L}} [ 1 + (h_{m_{L}}-1)\epsilon_{L}] + O(\epsilon_{L}),
\end{equation}
\begin{equation}
I_{3}^{'} = \frac{-1}{8 \epsilon_{L}}
[1+(2h_{m_{L}} - \frac{5}{4}) \epsilon_{L}] - \frac{1}{8}I 
+ O(\epsilon_{L}), 
\end{equation}
\begin{equation}
I_{4} = \frac{1}{2 \epsilon_{L}^{2}} \Bigl(1 +
(2\;h_{m_{L}}  -\frac{3}{2}) \epsilon_{L} + O(\epsilon_{L}^{2})\Bigr),
\end{equation}
\begin{equation}
I_{5}^{'} =
\frac{-1}{6 \epsilon_{L}^{2}}
[1+(3h_{m_{L}} -\frac{7}{4})\epsilon_{L}+ O(\epsilon_{L}^{2} ] - \frac{1}{4\epsilon_{L}} I,
\end{equation}
where 
$h_{m_{L}} = 1 + \frac{(\psi(1) - \psi(2- \sum_{n=2}^{L}\frac{m_{n}}{2n}))}
{2}$ and 
\begin{equation}
I= \int_{0}^{1}dx (\frac{1}{1-x(1-x)} + \frac{ln[x(1-x)]}{[1-x(1-x)]^{2}}) ,
\end{equation}
is the same integral occurring in the original work by BLZ and in the 
$m$-axial Lifshitz anisotropic integrals. This integral drops out in the 
calculation of the critical exponents by extensive cancellations in the 
renormalization functions at the fixed point. As pointed out in the previous 
massive approach for the $m$-fold anisotropic Lifshitz criticality, its 
appearing is a peculiarity of the orthogonal approximation. The normalization 
conditions and our choice of the subtraction point are responsible for this 
nonuniversal function, since it does not show up in the 
$\epsilon_{L}$-expansion of these integrals in the massless case.

\section{Isotropic graphs in the generalized orthogonal approximation}
We shall pursue the analogy with the anisotropic case in order to compute 
the isotropic integrals $(d=m_{n})$ in the orthogonal approximation. There 
is only one 
type of subspace(/competition axes) to be integrated over and the parameter 
is now $\epsilon_{n}=4n-d$. The minimal set of loop integrals now read: 
\begin{equation}
I_2 =  \int \frac{d^{m_{n}}k}{\bigl((k^{2})^{n} + 1 \bigr)^{2}}\;\;\;,
\end{equation}

\begin{equation}
I_{3} = \int \frac{d^{m_{n}}k_{1} d^{m_{n}}k_{2}}
{\bigl(((k_{1} + k_{2} + K)^{2})^{n} + 1 \bigr)[(k_{1}^{2})^{n}+1][(k_{2}^{2})^{n}+1]}\;\;\;,
\end{equation}

\begin{eqnarray}
I_{4}\;\; =&& \int \frac{d^{m_{n}}k_{1}d^{m_{n}}k_{2}}
{[(k_{1}^{2})^{n}+1]^{2} 
\left((k_{2}^{2})^{n} + 1 \right) \bigl(((k_{1} + k_{2})^{2})^{n} + 1 \bigl)}\;\;,
\end{eqnarray}

\begin{equation}
I_{5} = \int \frac{d^{m_{n}}k_{1} d^{m_{n}}k_{2}d^{m_{n}}k_{3}}
{\bigl(((k_{1} + k_{2} + K)^{2})^{n} + 1 \bigr) \bigl(((k_{1} + k_{3} + K)^{2})^{n} + 1 \bigr)((k_{1}^{2})^{n} + 1)  ((k_{2}^{2})^{n}+ 1) ((k_{3}^{2})^{n} + 1) }\;\;\;.
\end{equation}
One should keep in mind that the derivatives with respect to $K^{2n}$ of the 
two-point functions, namely  $I_{3}^{'}$ and $I_{5}^{'}$, are actually the 
objects of interest in our discussion. Recall that these integrals are 
evaluated at $K=0$. Therefore, using the same technology as before, the calculation is 
even simpler than in the anisotropic case. Again, every time a loop integral 
is performed an angular factor takes place. The geometric angular factor to be 
absorbed in a redefinition of the coupling constant in the isotropic integrals 
is $S_{m_{n}}$, which corresponds to the area of the unit sphere in $d=m_{n}$ 
dimensions. Omitting this factor, we encounter the following results 
for the required integrals:
\begin{equation}
I_{2} = \frac{1}{\epsilon_{n}} [ 1 - \frac{1}{2n} \epsilon_{n}] + O(\epsilon_{n}),
\end{equation}
\begin{equation}
I_{3}^{'} = \frac{-1}{8n \epsilon_{n}}
[1 - \frac{1}{4n} \epsilon_{n}] - \frac{1}{8 n^{2}}I 
+ O(\epsilon_{n}), 
\end{equation}
\begin{equation}
I_{4} = \frac{1}{2 \epsilon_{n}^{2}} \Bigl(1 - \frac{1}{2n}\epsilon_{n} + O(\epsilon_{n}^{2})\Bigr),
\end{equation}
\begin{equation}
I_{5}^{'} =
\frac{-1}{6n \epsilon_{n}^{2}}
[1 - \frac{1}{4n}\epsilon_{n}+ O(\epsilon_{n}^{2}) ] - \frac{1}{4 n^{2}\epsilon_{n}} I,
\end{equation}
where $I$ is the same integral taking place in Appendix A. Similarly to 
the situation found in the anisotropic cases, it does not contribute to the 
isotropic critical exponents as well.

\end{document}